\documentclass[11pt,prd,superscriptaddress,nofootinbib]{revtex4-2}
\usepackage{graphicx}
\usepackage{float}
\usepackage{placeins}
\usepackage[colorlinks=true, citecolor=blue, linkcolor=blue, urlcolor=blue]{hyperref}
\usepackage[dvipsnames]{xcolor}
\usepackage{booktabs}
\usepackage[
    colorlinks=true,
    linkcolor=blue,
    citecolor=blue,
    urlcolor=blue
]{hyperref}

\usepackage{amsmath}

\usepackage[colorlinks=true, linkcolor=blue, citecolor=blue, urlcolor=blue]{hyperref}

\begin{document}

\title{$N \to \Delta$ Transition Form Factors in the Thermal AdS/QCD Soft-Wall Model with Modified Propagators} 

\author{Narmin Nasibova\footnote{These authors contributed equally to this work.}} 
\email{n.nesibli88@gmail.com}
\affiliation{Institute of Physics, Ministry of Science and Education, H. Javid Avenue 131, AZ1073, Baku, Azerbaijan}
\affiliation{Center for Theoretical Physics, Khazar University, 41 Mehseti Str., AZ1096, Baku, Azerbaijan}
\affiliation{Baku Engineering University, Khirdalan city, Hasan Aliyev Str., 120 AZ0101, Baku, Azerbaijan}

\author{Xerxes D. Arsiwalla\footnotemark[1]} 
\email{x.d.arsiwalla@gmail.com}
\affiliation{Wolfram Institute for Computational Foundations of Science, Champaign, IL, USA}

\author{Shakir Nagiyev}
\email{shakir.m.nagiyev@gmail.com}
\affiliation{Institute of Physics, Ministry of Science and Education, H. Javid Avenue 131, AZ1073, Baku, Azerbaijan}

\maketitle
\thispagestyle{empty}

\vspace{1cm}

\centerline{\bf Abstract} 
		
\vskip 0.5 cm
						
We investigate electromagnetic (EM) and axial vector (AV) form factors of the nucleon-Delta baryon transition within the thermal AdS/QCD soft-wall model. While the vector bulk-to-boundary propagator admits a well-behaved analytical solution, the transverse  axial vector equation of motion involves the vacuum expectation value of the bifundamental scalar (generating chiral symmetry breaking). Due to this, the axial propagator diverges asymptotically, rendering it unsuitable for computing observables in the soft-wall model. To address this problem, we implement a regularization for the axial propagator in the form of a momentum-dependent cut-off, that identifies its correct physical regime. This is achieved via an interior constraint approach to the boundary-value problem, that determines the cut-off point. The consistency of this procedure is verified near the UV limit. As a proof of concept, we test this modified  propagator (and also a parametrically deformed version of it) to estimate axial vector form factors of the $N \to \Delta$ transition, and find results moderately  comparable with form factor profiles from experimental observations, lattice QCD, and other theoretical models. Equipped with this regularization, the thermal soft-wall model becomes a useful tool to probe thermal characteristics of QCD form factors. We demonstrate this for both, the EM and AV sectors of the $N \to \Delta$ transition.




\tableofcontents

\vspace{1cm}
\section{Introduction}

The Anti-de Sitter/Conformal Field Theory (AdS/CFT) correspondence \cite{maldacena1999large, witten1998anti, witten2001anti,  gubser1998gauge, nastase2007introduction}, or holographic duality, continues to be a useful tool to study non-perturbative aspects of  quantum field theories. In this set-up, the gravitational field lives in the  5-dimensional curved AdS  bulk, which is a solution of Einstein's field equations with a negative cosmological constant. The conformal field theory (CFT) describes the physics on the AdS boundary, constituting the 4-dimensional flat Minkowski spacetime. Using holography as a tool, one can now compute observables in QCD-like theories, which lie beyond the realm of perturbative techniques. The most widely used realizations of holographic QCD are the hard-wall and the soft-wall model. These models implement confinement by modifying the infrared behavior of the AdS background, either through the introduction of a sharp cutoff (hard-wall) \cite{erlich2005qcd, da2005chiral, hong2006couplings}, or via a smooth dilaton profile (soft-wall) \cite{karch2006linear}. Both have been extensively applied in high-energy physics to make phenomenological predictions in QCD, particularly,  nonperturbative observables of strong interactions  \cite{gubser1998gauge, colangelo2012medium, mamani2014vector, braga2017holographic,  bartz2016chiral, vega2017analysis, afonin2016soft, chen2019critical}. Furthermore, in \cite{gutsche2019mesons, gutsche2019baryons}, the derivation of analytical formulas for the mass spectra of mesons and baryons, and their coupling constants were studied at low temperatures; and in   \cite{gutsche2020electromagnetic}, the electromagnetic (EM) form factors of nucleons were  investigated in the soft-wall model of AdS/QCD by taking into account the temperature of the medium: the so-called thermal soft-wall model. 

As further evidence in favor of the thermal soft-wall model, in \cite{mamedov2021temperature}, the temperature dependence of the vector meson-nucleon coupling constant has been studied. In \cite{nasibova2021temperature, nasibova2024isospin, nasibova2022meson, nasibova2026pion}, it was found that the coupling constants $g_{\rho NN} (T)$, $g_{\rho\Delta\Delta}(T)$ and $g_{\omega NN}(T)$ decrease with increasing temperature, and they vanish at the confinement-deconfinement temperature, which corresponds to the Hawking temperature in the gravitational theory. Furthermore, in \cite{nasibova2025pion}, the pion form factor, its GPD, radius, mass and  charge density \cite{nasibova2026pion} were investigated at finite temperature using the thermal soft-wall model, with results corroborating with known estimates. These studies, at the least, suggest the utility of holographic methods for QCD phenomenology, and in particular, the use of the thermal soft-wall model.


One of our objectives in this work is to investigate the EM and axial vector form factors of the $N \to \Delta$ transition within the thermal soft-wall model of AdS/QCD. The other objective (mentioned below) is to address the divergence in the axial propagator. Now, the  $N\to\Delta$ transition is phenomenologically important because it simultaneously encodes hadron deformation, chiral dynamics, and the weak axial response that drives $\Delta$ production in neutrino reactions in hadronic matter at finite temperatures. Furthermore, its thermal evolution offers a window into how the spin-isospin structure of baryons is reshaped on the way to deconfinement. Hence, this transition is significant for astrophysical phenomena as well as heavy-ion collisions. On the other hand, testing models with this transition serves as a stringent theoretical benchmark due to the complex nature of the multipole form factors involved with it. In fact, the  \emph{complete characterization} (and not approximations or limiting cases) of, particularly, the axial vector form factors of the $N \to \Delta$ transition have not been addressed so far within the soft-wall model. The troublesome   issue involves the axial vector bulk-to-boundary propagator, which is obtained by solving the EOM of the transversal axial gauge field. This EOM involves a 5D  Higgs-like mass term for the axial  field, generated by the chiral symmetry breaking vacuum of the bifundamental scalar. By a complete characterization of the axial propagator, we mean a solution to this EOM in the presence of the mentioned VEV term. Implementing this solution has not been possible in the soft-wall model.   

The reason for the above problem is that the full numerical solution for the axial vector propagator diverges  asymptotically. More recently, this and other such divergences have been systematically studied in \cite{leutgeb2026divergences}. For this reason, in earlier work   \cite{mamedov2023axial}, approximations were used in place of the full solution, namely, the bulk solution of the axial propagator was taken to be of the same functional form as the vector EM propagator (a similar approximation has also been used by several authors working with the hard-wall model \cite{maru2009meson, mamo2021neutrino}). This is a reasonable assumption when chiral symmetry holds, but too coarse-grained, an approximation, beyond that. In \cite{leutgeb2026divergences}, it was suggested that such a divergence in the propagator is indicative of the fact that the soft-wall model itself requires a modification. 

To resolve the issue caused by the divergence in the axial propagator, here we propose a  modification that still leaves the core of the model (Lagrangian, interactions, Regge trajectories, etc.) untouched. We implement a regularization for the propagator in the form of a momentum-dependent cut-off, that identifies its physical regime. For this, we solve the standard axial vector EOM numerically, as a boundary-value problem with an interior constraint, and then implement a cut-off similar to (but not the same as) the hard-wall model. 
As a proof of concept, we then proceed to test this modified propagator to estimate the axial vector form factors of the $N \to \Delta$ transition. 
Besides the standard axial EOM, we also test another modification of the propagator: namely, a parametrically deformed version of the axial EOM, where the sign of the VEV term is reversed. Asymptotically, this also diverges, and has to undergo the same regularization procedure.  Interestingly, the cut-off propagator constructed using the deformed EOM yields results compatible with known (from lattice QCD studies) axial form factor profiles, including their thermal evolution. On the other hand, the form factor estimates we get using the standard axial vector propagator are comparable to those obtained using the AdS/QCD hard-wall model \cite{ahn2009spin}.  Our modified propagators thus provides a plausible work-around for the axial sector of the soft-wall model, and produces physically meaningful results. Beyond the axial sector, more work will be needed to determine whether similar regularization procedures may help circumnavigate other divergences reported in \cite{leutgeb2026divergences}; though, that is outside the scope of our work here.



 The remainder of this paper is organized as follows:  \autoref{soft} and \autoref{va} respectively concern the holographic soft-wall model, and the correct treatment of the vector and axial vector bulk-to-boundary propagators ($V(Q,r, T)$ and $A(Q,r, T)$) at finite temperature. In  \autoref{ff},  
 we construct the temperature-dependent integral expression for the $EM$ and $AV$ form factors of the $N \to \Delta$ transition.    
 In \autoref{nu}, we present our numerical results, including the temperature dependence of the $EM$ and $AV$ form factors, validating them against experimental measurements or other theoretical approaches, where applicable. Finally, in \autoref{con}, we conclude with discussions.

\section{Soft-Wall Model at Zero and Finite Temperatures}

\label{soft}

The action in the ordinary soft-wall model is written as follows:
\begin{equation}
S=\int d^{4}x dz\sqrt{g}e^{-\varphi(z)}L(x,z),
\label{3a}
\end{equation}
where $g=|det \, g_{M N}|$ is the determinant of the AdS metric. Here $(M, N=0,1,2,3,5)$. And $z$ is the holographic coordinate, which varies throughout the interval $0 \leq z < \infty$. $x=(t,\vec x)$ is the set of Minkowski coordinates. $L(x,z)$ is the Lagrangian that represents the 5-dimensional interactions between bulk fields. $\varphi(z)$ is the dilaton field in the soft-wall model, which is responsible for breaking the chiral symmetry with an exponential factor, so that the integral over the $z$ coordinate becomes finite at the IR boundary $(z\rightarrow \infty )$. 
The expression for the quadratic dilaton field in the soft-wall model is as follows:
	\begin{equation}
		\varphi(z)=k^{2}z^{2}
		\label{a}
	\end{equation}
    Here,  $k$ is the scale parameter with the value of a few hundred $MeV$ \cite{gutsche2019mesons, gutsche2019baryons}.  
The background geometry is assumed to be 5D AdS spacetime with the AdS metric at the zero-temperature limit in the following form:  
	\begin{equation}
		ds^{2}=e^{2A(z)}\left[dt^{2}-d{\vec{x}}^{2}-dz^{2}
        \right].\label{5a}
	\end{equation}
	Here, the function  $A(r)=log(\frac{R}{r})$, where $R=1$ is the AdS radius.

The action in the thermal soft-wall model with the temperature-dependent dilaton field has the following form \cite{gutsche2019mesons,gutsche2019baryons,gutsche2020electromagnetic,mamedov2021temperature,mamedov2023axial,nasibova2024isospin,nasibova2025pion}:
\begin{equation}
S=\int d^{4}x dr\sqrt{g}e^{-\varphi(r,T)}L(x,r,T),
\label{3}
\end{equation}
Here, $r$  is the Regge-Wheeler tortoise  coordinate, which is used instead of the $z$ coordinate in the thermal soft-wall. The relationship between these two coordinates is as follows \cite{regge1957stability,horowitz2000quasinormal}:
\begin{equation}
r=\int \frac{dz}{f(z)}
\label{1}
\end{equation}
Moreover, the $r$ coordinate can be
expanded as (in terms of the $z$ coordinate):  
\begin{equation}
r\approx z\left[1+\frac{z^{4}}{5z_{H}^{4}}+\frac{z^{8}}{9z_{H}^{8}}\right].
\label{2}
\end{equation}
At high temperatures, the metric is the AdS Schwarzschild metric, which takes the form:  
	\begin{equation}
ds^{2}=e^{2A(r)}\left[f(r,T)dt^{2}-d{\vec{x}}^{2}-\frac{dr^{2}}{f(r,T)}\right].\label{5}
	\end{equation}
	Here $f(r,T)$ is the thermal factor, which is expressed as: \cite{gutsche2019mesons,gutsche2019baryons,nasibova2022meson}:
	\begin{equation}
		f(r,T)=1-\frac{r^{4}}{r_{H}^{4}}, \label{6}
	\end{equation}
	where $r_{H}$ is the position of the event horizon and is related to the Hawking temperature as $T = 1/(\pi \, r_{H})$.  At low temperatures $f(r,T)=1$ (which is the case that we focus on) it will turn to the Poincare metric. 
    
In the thermal action (Eq. \ref{3}), $\varphi(r, T)$ is the thermal dilaton field which attains a temperature dependence of the following form:  
\begin{equation}
		\varphi(r,T)=K^{2}(T)r^{2}
		\label{4}
	\end{equation}
where the parameter $K^{2}(T)$ represents the spontaneous breaking of chiral symmetry at finite temperatures. Its explicit form has been established  \cite{gutsche2019baryons,gutsche2019mesons}:
	\begin{equation}
		K^{2}(T)=k^{2}[1+\rho(T)+O(T^{6})].
		\label{5}
	\end{equation}
    The relation between $K^{2}(T)$ and $\rho(T)$ comes from the low-temperature dependence of  quark condensates using two-loop chiral perturbation theory (ChPT) \cite{gasser1987light}. 
 The $T$ dependence of the dilaton field is encoded in the $\rho(T)$ function which was found to be \cite{gutsche2019baryons,gutsche2019mesons}:
	\begin{equation} 
		\rho(T)=\frac{9\alpha\pi^{2}}{16}\frac{T^{2}}{12F^{2}}-\frac{ N_{f}^{2}-1}{N_{f}}\frac{T^{2}}{12F^{2}}-\frac{N_{f}^{2}-1}{2N_{f}^{2}}\left(\frac{T^2}{12F^2}\right)^{2} +O\left(T^6\right).
		\label{13}
	\end{equation}
In this case, the number of quark flavors is $N_{f}=2$. The  pion decay constant $F$ is related to the $k$ parameter as $F=\frac{k\sqrt{6}}{8}$ and $\alpha$ is the parameter  encoding the gravitational contribution   \cite{gutsche2019mesons}.   

\section{Vector and Axial Vector Propagators}
\label{va}
In the thermal soft-wall model, the five-dimensional action for the vector field $V_M^a(x,r,T)$ and the axial vector field $A_M^a(x,r,T)$ is as follows:
\begin{equation}
S = -\frac{1}{4 g_5^2} \int d^5 x \, \sqrt{g} \, e^{-\varphi(r,T)}
\left(
F_V^{MN} F^V_{MN}
+
F_A^{MN} F^A_{MN}
\right),
\label{act}
\end{equation}
where the field strength tensors are defined as 
$F_V^{MN} = \partial^M V^N -\partial^N V^M$ for the vector field, and for the axial  vector field is 
$F_A^{MN}= \partial^M A^N - \partial^N A^M - i [A^M, A^N]$, 
with the latter corresponding to the non-Abelian axial vector field. 
$g_5$ denotes the five-dimensional gauge coupling. 
In the quadratic approximation and in the Lorentz gauge, the equation of motion (EOM) for the vector field is obtained from the  action (Eq. \ref{act}) in the thermal soft-wall model:  \cite{gutsche2019mesons, nasibova2025pion}:
\begin{equation}
    \partial_r \left( \frac{e^{-\varphi(r,T)}}{r} \partial_rV(Q,r,T) \right) - Q^2 \frac{e^{-\varphi(r,T)}}{r} V(Q,r,T) = 0, 
    \label{vk}
    \end{equation}
    And, the EOM (equation of motion) for the axial field is as follows:
    \begin{equation}
    \partial_r \left( \frac{e^{-\varphi(r,T)}}{r} \partial_r A(Q,r,T) \right) - Q^2 \, \frac{e^{-\varphi(r, T)}}{r} A(Q,r,T) - g_5^2 \, \frac{v^2(r, T) \, e^{-\varphi(r,T)}}{r^3} A(Q,r,T) = 0,
    \label{ax}
 \end{equation}
Here, $V(Q,r,T)$ and $A(Q,r,T)$ denote the bulk-to-boundary propagators for the vector  and axial vector fields, respectively, corresponding to a four-dimensional momentum $Q$. And $v(r, T)$ is the chiral symmetry breaking vacuum expectation value relevant only for the axial vector field.  

In fact, Eq.~\ref{ax} can be written more generally with a parametric deformation as  
\begin{equation}
    \partial_r \left( \frac{e^{-\varphi(r,T)}}{r} \partial_r A(Q,r,T) \right) - Q^2 \, \frac{e^{-\varphi(r, T)}}{r} A(Q,r,T) - \tau  \, \frac{ g_5^2 \, v^2(r, T) \, e^{-\varphi(r,T)}}{r^3} A(Q,r,T) = 0,
    \label{ax2}
 \end{equation}
where $\tau$ is the deformation parameter. When $\tau = 1$, we get the axial EOM (Eq.~\ref{ax}), and when $\tau = 0$, we get the vector EOM (Eq.~\ref{vk}). This deformation interpolates across vector and axial sectors. It may be interesting to investigate whether such a continuous deformation might have phenomenological significance. For now, besides $\tau$ being $0$ and $1$, we shall consider only one other deformation, when $\tau = -1$, for reasons that shall become clear in what follows. 
    
Returning to the EOM for the vector propagator (Eq.~\ref{vk}), the   boundary condition for the vector field is taken to be $V(Q,0,T) =1$ \cite{gutsche2019mesons}. This leads to the solution  \cite{gutsche2019baryons, gutsche2019mesons}:
  \begin{equation}
		V(Q, r, T)=\Gamma \left(1+a(Q, T)\right)U\left(a(Q, T), 0, K^{2}r^{2}\right)=K^{2}r^{2}\int_{0}^{1}\frac{dx}{(1-x)^{2}}x^{a(Q, T)}e^{-K^{2}r^{2}\frac{x}{1-x}},
		\label{15}
	\end{equation}
	Here $\Gamma$ is Euler's gamma function and $a(Q,T)=\frac{Q^{2}}{4K^{2}(T)}$. And $U(x,y,z)$ is a Tricomi function, also known as the confluent hypergeometric function of the second kind. 

In the holographic approach to QCD, the axial vector sector serves as a crucial probe for both confinement and  chiral symmetry breaking. We consider the axial vector bulk-to-boundary propagator, $A(Q,r,T)$, which mediates the interaction between  4D boundary current sources and the 5D bulk gauge fields in a hot medium. In the soft-wall background, the dynamics are governed by the action in (Eq.~\ref{act})  and the corresponding EOM (Eq.~\ref{ax}), which can alternatively be expressed in a standard second-order form as follows: 
\begin{equation}
A''(Q,r,T) - \left( 2 K^2(T) \, r + \frac{1}{r} \right) A'(Q,r,T) 
- \left( Q^2 + \frac{g_5^2 \; v^2(r,T)}{r^2} \right) A(Q,r,T) = 0,
\label{l}
\end{equation}
The function  $v(r,T)$ has a solution of the form \cite{gutsche2020electromagnetic}: 
	\begin{equation}
		v(r,T)=\frac{1}{2}M_{q}  \, r + \frac{1}{2}\Sigma(T) \, r^{3}. 
		\label{32}
	\end{equation}
where, the parameters $M_{q}$ and  $\Sigma$ are identified with the $u$/$d$ quark mass, and with the chiral condensate $\Sigma (T)= \langle 0|\bar{q}q|0 \rangle _T$, respectively.   
The function $\Sigma(T)$, it was found in  \cite{gasser1987light} by using two-loop chiral perturbation theory at finite temperature, and applied in  \cite{gutsche2019mesons,gutsche2019baryons} for the finite-temperature soft-wall model.  
In \cite{da2006scalar, karch2006linear}, the EOM for the axial vector field  coincides with the EOM for the vector field when $(m_q\rightarrow 0, z\rightarrow z_{UV})$. 
In both the hard-wall and soft-wall AdS/QCD frameworks, the Yukawa-type
interaction term, which is proportional to the scalar vacuum expectation
value $v(z)$, is omitted in deriving and solving the fermionic equations of motion. This simplification follows the standard treatment adopted in  
\cite{maru2009meson,gutsche2019mesons,gutsche2019baryons}.
In \cite{maru2009meson}, $v(z)$ was also not considered when solving the EOM  for the bulk axial field at zero temperature. 
In \cite{mamedov2023axial}, 
by taking $v(r, T)=0$ the EOM, it was solved, and the axial vector propagator was taken to be of the same form as the vector propagator in the thermal soft-wall model.
In our work here, we will work on a more general approach to solve for the axial vector propagator, which does not require one to enforce $v(r, T)=0$. We will work with the full numerical solution for axial vector propagator at finite temperature.

\subsection{Boundary-Value Problem \& Regularization for the Axial Vector Propagator}

In the soft-wall model, unlike the vector boson, the equation of motion (EOM) for the axial vector propagator (Eq.~(\ref{l})) has a  potential sourced by the bulk scalar VEV. The vector boson propagator can, in fact, be solved analytically, in terms of the confluent hypergeometric function (Eq.~(\ref{15})). However, the axial vector EOM is akin to what is known as an ``extended confluent hypergeometric differential equation" (the axial equation is actually a bit more general than that because it also includes negative powers of $r$) \cite{campos2001some}. As has been recently reported, the problem with the axial bulk-to-boundary propagator is that it is divergent \cite{leutgeb2026divergences}. More specifically, the solution to the aforementioned EOM leads to an asymptotic divergence. This is particularly problematic for the soft-wall model which requires integrating the propagator through the full range of the holographic coordinate, when computing observables. On the other hand, the hard-wall model evades this issue by implementing a rigid IR cut-off for all observables. A possible resolution to this problem within the soft-wall model would be to determine a ``soft" IR cut-off for the axial propagator based on consistency conditions, rather than a rigidly imposed IR boundary condition. This is what we will do to regularize the axial propagator. 
 
We first determine the appropriate boundary conditions required for physically well-defined solutions. At $r = 0$, we choose $A(Q, 0, T) = 1$, which ensures a non-zero, finite mode at the origin. The second boundary condition is the one that requires careful treatment. As mentioned, the solution is divergent at $r \rightarrow \infty$ (which was checked numerically). Therefore, we will work with a boundary-value problem with an interior constraint. The latter shall be determined by examining conditions near the UV limit, where the axial vector EOM approaches the form of the EOM for the vector boson propagator, up to a constant shift. Given that the  solution for the vector propagator is known analytically, we will use that to find the second boundary condition of the axial EOM. Notice that the additional terms in the potential of the axial vector EOM (compared to the vector EOM) are 
\[  g_{5}^{2}  \left( \frac{1}{2}M_{q}  + \frac{1}{2}\Sigma(T) \,  r^{2}\right)^{2} \] 
As $r \rightarrow 0$, this potential converges to a constant, proportional to $g_5^2 \, M_q^2$. On the other hand, when $r \gg 1$, this gives a quartic $r^4$ contribution to the EOM. Therefore, for $r \ll 1$, and close to the UV limit, this potential has a contribution of order ${\cal O}(r^0)$, namely, $- (Q^2 + {\cal O}(r^0))$. In that region of $r$, the axial EOM approaches the same form as the vector  EOM (Eq.~(\ref{vk})) up to a shift  
\[ Q^2 \to Q^2 + {\cal O}(r^0) \]
Using the analytic solution for the vector propagator $V (Q, r, T)$
(Eq.~(\ref{15})), our second boundary condition (interior constraint) can therefore be taken at $r = r_0$ such that  
\begin{eqnarray}
    A(Q, \, r_0, \, T) \; = \;  V (Q + {\cal O}(r_0^0), \; r_0, \, T) 
\end{eqnarray} 
With this interior constraint, we can numerically solve the axial vector ODE using the standard Runge-Kutta method. For the current problem, we take  $r_{0} = 0.1$. As a check, this gives: 
\[  g_{5}^{2}  \left( \frac{1}{2}M_{q}  + \frac{1}{2}\Sigma(0) \,  r_0^{2}\right)^{2}  \approx  0.000076    \]
at $T=0$ (and decreases even further for $T > 0$), which is consistent with the method. In this way, we obtain numerical values for $A (Q, r, T)$, over a range of different $Q$'s and $T$'s. 

So far, what we have discussed above serves as a means to consistently determine the second boundary condition, given that the solution grows rapidly at large $r$. However, as mentioned earlier, the axial vector solution has divergent modes at $r \rightarrow \infty$. How should we handle those in such a way as to obtain physically meaningful solutions for the bulk-to-boundary propagator? This can be achieved by implementing an IR cut-off on $A (Q, r, T)$ at $r = r_{IR}$ whenever 
\[ \frac{\partial A (Q, r, T)}{\partial r} = 0  \] 
This is similar to the hard-wall model, but unlike the latter, here $r_{IR}$ is not a fixed cut-off because it is momentum and temperature dependent. Furthermore, to ensure that $A (Q, r, T)$ remains physically well-defined, we constrain the propagator to be bounded:   
\[ 0 \leq A (Q, r, T) \leq 1 \]  
Here, the upper bound (that is, $1$) is simply a way to impose normalization of form factors,  whereas the lower bound, which excludes negative values, is there to ensure that form factors such as  
\[ F(Q^2, T) = \int dr \, \phi^2 (r, T) \; A (Q, r, T) \]
always remain non-negative (consistent with empirical observations). 
In cases where the vanishing of the derivative occurs outside of the above bounds, then the bounds supersede the vanishing derivative constraint, and the cut-off is imposed at that point. 
Unlike the hard-wall model, the cut-off point is neither rigidly fixed nor tuned by parametric fits, but follows as a result of the interior boundary condition (which, in turn, is consistent with the potential in the near-UV region). Given that the cut-off point depends on $Q$ and $T$,  we solve for each value of $Q$ and $T$, and obtain a finite numerical function for the axial vector propagator in the presence of chiral symmetry breaking. As we show, in our numerical results below, this regularization procedure for the axial propagator turns out to be surprising effective in obtaining empirically comparable form factor profiles.

\section{Vector and Axial Vector  Form Factors of $\Delta$ Baryons  }
\label{ff}
Next, we look at baryon electromagnetic 
and axial vector transition form factors in four dimensions, which are extracted from 
the matrix elements of the vector and axial vector currents between 
$\Delta$ and nucleon states as follows  \cite{ahn2009spin}: 
\begin{equation}
\langle\Delta (p')|J_{\mu}^{EM}|N(p)\rangle =
\bar{u}^{\sigma}(p'){\cal{O}}^{(EM)}_{\sigma\mu}(p,p')u(p),
\label{a1}
\end{equation}
\begin{equation}
\langle\Delta (p')|A_{\mu}^{3}|N(p)\rangle =
\bar{u}^{\sigma}(p'){\cal{O}}^{(A)}_{\sigma\mu}(p,p')u(p),
\label{a2}
\end{equation}
 The general structure of the transition operator $O^{\sigma\mu}$ is 
constructed by imposing Lorentz covariance, current conservation, and 
the spin-$3/2$ constraints. As a result, the operator can be expressed 
in terms of three independent tensor structures built from $q_\mu$, 
$P_\mu$, and the metric tensor, ensuring $q^\mu O_{\sigma\mu}=0$.
These matrix elements are parametrized 
in terms of invariant form factors. Here, $u^{\sigma}(p')$ and $u(p)$ 
denote the Rarita--Schwinger spinor for the $\Delta$ baryon and the 
Dirac spinor for the nucleon, carrying momenta $p'$ and $p$, respectively.
The operators ${\cal O}^{(A)}_{\sigma\mu}$ and ${\cal O}^{(EM)}_{\sigma\mu}$ 
were constructed assuming CP invariance and were expanded using Lorentz invariance and current conservation (see \cite{ahn2009spin}). 
Within the thermal soft-wall model, we assume that the operator 
structures given in Eqs.~(\ref{a1}) and (\ref{a2}) acquire an explicit 
temperature dependence and can be written in the following form   \cite{alexandrou2007axial, Jones1973,ahn2009spin}:
\begin{eqnarray}
\label{EMff1}
{\cal O}^{(\mathrm{EM})}_{\sigma\mu}(p',p,T)
=\Big[
G_1(Q^2,T)\left(Q_{\sigma}\gamma_{\mu}
- Q\!\cdot\!\gamma\,\eta_{\sigma\mu}\right)
+G_2(Q^2,T)\left(Q_{\sigma}P_{\mu}
- Q\!\cdot\!P\,\eta_{\sigma\mu}\right)
\nonumber\\
+G_3(Q^2,T)\left(Q_{\sigma}Q_{\mu}
- Q^2\eta_{\sigma\mu}\right)
\Big]\gamma^5 .
\end{eqnarray}
 The axial vector transition operator is constructed by imposing Lorentz 
covariance and the spin-$3/2$ constraints, without requiring current 
conservation due to the partially conserved axial current. As a result, 
four independent structures appear, leading to the axial form factors 
$C_3$, $C_4$, $C_5$, and $C_6$.
\begin{eqnarray}
{\cal{O}}^{(A)}_{\sigma\mu}\!(p',p,T)=\!
\frac{C_3(Q^2,T)}{m_N}(Q\cdot\gamma\eta_{\sigma\mu}\!-\!Q_{\sigma}\gamma_{\mu})\!+\!\frac{C_4(Q^2,T)}{m_N^2}(p'\cdot
Q\eta_{\sigma\mu}\!-\!Q_{\sigma}p'_{\mu})\!\nonumber\\
+\!C_5(Q^2,T)\eta_{\sigma\mu}\!+\!\frac{C_6(Q^2,T)}{m_N^2}Q_{\sigma}Q_{\mu},
\label{Aff}
\end{eqnarray}
 Here, the kinematical variables are as 
$Q = p' - p$ and $P = (p' + p)/2$. 


The vertex operator for the electromagnetic matrix elements in Eq.~(\ref{EMff1}) can be also expressed in terms of the three Sachs form
factors corresponding to ~\cite{alexandrou2007axial,Jones1973} at finite temperature as follow:
\begin{equation}\label{EMff2}
{\cal{O}}^{(EM)}_{\sigma\mu}(p',p,T)=G_{M1}(Q^2,T)K^{M1}_{\sigma\mu}+G_{E2}(Q^2,T)K^{E2}_{\sigma\mu}
+G_{C2}(Q^2, T)K^{C2}_{\sigma\mu}.
\end{equation}
Here, $G_{M1}(Q^2,T)$, $G_{E2}(Q^2,T)$ and $G_{C2}(Q^2,T)$ are the magnetic dipole, electric
quadrupole and Coulomb quadrupole form factors respectively. The definitions of $K^{M1}_{\sigma\mu}$, $K^{E2}_{\sigma\mu}$, and $K^{C2}_{\sigma\mu}$ are  as follows \cite{ahn2009spin}:

\begin{equation}
K^{M1}_{\sigma\mu}=-\frac{3}{(m_{\Delta}+m_N)^2-Q^2}\frac{m_{\Delta}+m_N}{2m_N}
i\epsilon_{\sigma\mu\alpha\beta}p^{\alpha}p'^{\beta}, 
\end{equation}
\begin{equation}
K^{E2}_{\sigma\mu}=-K^{M1}_{\sigma\mu}+6\Omega^{-1}(Q^2)\frac{m_{\Delta}+m_N}{2m_N}
2i\gamma^5\epsilon_{\sigma\lambda\alpha\beta}p^{\alpha}p'^{\beta}{\epsilon_{\mu}}^{\lambda\gamma\delta}p_{\gamma}p'_{\delta},
\end{equation}
\begin{equation}
K^{C2}_{\sigma\mu}=-6\Omega^{-1}(Q^2)\frac{m_{\Delta}+m_N}{2m_N}i\gamma^5q_{\sigma}(Q^2(p+p')_{\mu}-q\cdot
(p+p')q_{\mu}),
\end{equation}
where
\begin{equation}
\Omega(Q^2)=\left[(m_{\Delta}+m_N)^2-Q^2\right]\left[(m_{\Delta}-m_N)^2-Q^2\right].
\end{equation}
The magnetic dipole ($G_{M_{1}}(Q^2,T)$) , electric quadrupole ($G_{E_{2}}(Q^2,T)$), and Coulomb quadrupole ($G_{C_{2}}(Q^2,T)$ transition 
form factors respectively are written in terms of the invariant form factors $G_i(Q^2,T)$ 
as formulated in .~\cite{Jones1973,ahn2009spin}. 
The transition current between the nucleon and the $\Delta$ baryon can be  
expressed in terms of three independent Lorentz structures, such as 
$Q_\sigma \gamma_\mu$, $Q_\sigma P_\mu$, and $Q_\sigma Q_\mu$, which 
lead to the invariant form factors $G_1$, $G_2$, and $G_3$. 
These structures provide a complete and Lorentz-covariant parametrization 
of the spin-$1/2$ $\to$   spin-$3/2$ transition current.

However, the invariant form factors do not have a direct physical interpretation, 
as the corresponding Lorentz structures mix different multipole contributions. 
For this reason, it is convenient to introduce the Jones–Scadron transition multipole form factors 
$G_M$, $G_E$, and $G_C$, which correspond to the magnetic dipole, 
electric quadrupole, and Coulomb quadrupole transitions, respectively \cite{Jones1973}.

The Jones–Scadron form factors are related to the invariant form factors through 
linear combinations, effectively representing a change of basis that separates 
the magnetic and electric contributions and allows for a direct comparison 
with experimental observables.
Introducing  
$\hat m \equiv \frac{m_N}{m_\Delta + m_N}$, we obtain at finite temperature
\begin{align}
G_{M_1}(Q^2,T) ={}& \Big[ \frac{(3m_{\Delta} + m_N)(m_{\Delta} + m_N) + Q^2}{m_{\Delta}} G_1(Q^2,T) \nonumber \\
& + (m_{\Delta}^2 - m_N^2) G_2(Q^2,T) - 2Q^2 G_3(Q^2,T) \Big] \frac{\hat{m}}{3}, \\[1.5ex]
G_{E_2}(Q^2,T) ={}& \Big[ \frac{(m_{\Delta}^2 - m_N^2) - Q^2}{m_{\Delta}} G_1(Q^2,T) \nonumber \\
& + (m_{\Delta}^2 - m_N^2) G_2(Q^2,T) - 2Q^2 G_3(Q^2,T) \Big] \frac{\hat{m}}{3}, \\[1.5ex]
G_{C_2}(Q^2,T) ={}& \Big[ 2m_{\Delta} G_1(Q^2,T) + \frac{1}{2}(3m_{\Delta}^2 + m_N^2 + Q^2) G_2(Q^2,T) \nonumber \\
& + (m_{\Delta}^2 - m_N^2 - Q^2) G_3(Q^2,T) \Big] \frac{2\hat{m}}{3}.
\end{align}

The ratios of the electric and Coulomb quadrupole amplitudes,  to the
magnetic dipole amplitude at finite temperature, $R_{EM}(T)$ (EMR) and $R_{SM}(T)$ (CMR), respectively, are 
defined  as
\begin{equation}
R_{EM}(T)=-\frac{G_{E_2}(Q^2,T)}{G_{M_1}(Q^2,T)}, \qquad\qquad
R_{SM}=-\frac{|\vec{q}|}{2m_{\Delta}}\frac{G_{C_2}(Q^2,T)}{G_{M_1}(Q^2,T)}.
\end{equation}
Here, $|\vec q|$ is the magnitude of the virtual photon four-momentum.
The thermal nucleon-$\Delta$ transition  EM and axial vector form factors are then obtained using the following interaction Lagrangian:  
\begin{eqnarray}
\mathcal{L}_{FN\Delta} &=& \alpha_1 \, \bar{\Psi}^M_1 \Gamma^N (F_L)_{MN} N_1 
+ i \alpha_2 \, (\partial^M \bar{\Psi}^N_1) (F_L)_{MN} N_1 
 + i \alpha_3 \, \bar{\Psi}^M_1 (\partial^N (F_L)_{MN}) N_1  \nonumber\\&&  + \beta_1 \, \bar{\Psi}^M_1 \Gamma^N \Gamma^P (\tilde{F}_L)_{MNP} N_1  + i \beta_2 \, (\partial^M \bar{\Psi}^N_1) \Gamma^P (\tilde{F}_L)_{MNP} N_1 + \tilde{\alpha}_1 \, \bar{\Psi}^M_1 \Gamma^N (F_L)_{MN} X N_2 \nonumber  \nonumber \\
&& +i \tilde{\alpha}_2 \, (\partial^M \bar{\Psi}^N_1) (F_L)_{MN} X N_2 \nonumber + i \tilde{\alpha}_3 \, \bar{\Psi}^M_1 (\partial^N (F_L)_{MN}) X N_2 + \tilde{\beta}_1 \, \bar{\Psi}^M_1 \Gamma^N \Gamma^P (\tilde{F}_L)_{MNP} X N_2 \nonumber \\
&& + i \tilde{\beta}_2 \, (\partial^M \bar{\Psi}^N_1) \Gamma^P (\tilde{F}_L)_{MNP} X N_2 + \text{h.c.}
\label{lag}
\end{eqnarray}
which involve 5-dimensional AdS bulk fields: namely, the nucleon $N_i$, delta baryon $\Psi_i$, scalar $X$, vector and axial vector gauge fields $F_L$. In the expression of the action (Eq.~\ref{3}) the  \(\alpha\) and \(\beta\) are unknown parameters in the model.  
After taking into account the interaction Lagrangian (Eq.~\ref{lag}) in the 5-dimensional action (Eq.~\ref{3}), the expressions for the invariant vector form factors of the nucleon to
$\Delta$ transition are obtained in the following form :   
\begin{eqnarray}
G_1(Q^2,T) = \int dr \frac{V(Q,r,T)}{r^2} \bigg\{ \!\! &\kappa& \!\! \Big( F^{\ast}_{1L}(p^{\prime},r,T)f_{1L}(p,r,T) - F^{\ast}_{2L}(p^{\prime},r,T)f_{2L}(p,r,T) \Big) \nonumber \\
&+& 2m_{\Delta}r \beta_2 \Big( F^{\ast}_{1L}(p^{\prime},r,T)f_{1R}(p,r,T) + F^{\ast}_{2L}(p^{\prime},r,T)f_{2R}(p,r,T) \Big) \nonumber \\
&+& 2r \beta_2 \Big( (\partial_r F^{\ast}_{1L}(p^{\prime},r,T ))f_{1L}(p,r,T) + (\partial_r F^{\ast}_{2L}(p^{\prime},r,T ))f_{2L}(p,r,T) \Big) \bigg\} \hspace{2em}  
\end{eqnarray}
\begin{eqnarray}
G_2(Q^2,T) &=& -\int \frac{dr}{r} V(Q,r,T) (\alpha_2 + 2\beta_2) \Big[ F^{\ast}_{1L}(p^{\prime},r,T)f_{1R}(p,r,T) + F^{\ast}_{2L}(p^{\prime},r,T)f_{2R}(p,r,T) \Big]  \\ [5.5ex]   
G_3(Q^2,T) &=& -\int \frac{dr}{2r} V(Q,r,T) (\alpha_2 + 2\beta_2 + 2\alpha_3) \Big[ F^{\ast}_{1L}(p^{\prime},r,T)f_{1R}(p,r,T) + F^{\ast}_{2L}(p^{\prime},r,T)f_{2R}(p,r,T) \Big]  \hspace{2em} 
\end{eqnarray}

where $F_{n,L/R}$ and $f_{n, L/R}$ are the nucleon and delta baryon profile functions respectively (see eqs.~\ref{ff1} and \ref{ff2} for their explicit form).
In the same way, the axial vector form factors for the  nucleon to
$\Delta$ transition in the Adler parametrization \cite{adler1968photo, schreiner1973neutrino}  are obtained in the following form \cite{ahn2009spin}:

\begin{align}
C_3(Q^2,T) ={}& -m_N \int dr \frac{A(Q,r,T)}{r^2} \bigg\{ 
\kappa \Big( F^{\ast}_{1L}(r,T)f_{1L}(r,T) + F^{\ast}_{2L}(r,T)f_{2L}(r,T) \Big) \nonumber \\
& + \tilde{\kappa}v(r,T) \Big( F^{\ast}_{1L}(r,T)f_{2L}(r,T) - F^{\ast}_{2L}(r,T)f_{1L}(r,T) \Big) \nonumber \\
& - 2m_{\Delta}r \beta_2 \Big( F^{\ast}_{1L}(r,T)f_{1R}(r,T) - F^{\ast}_{2L}(r,T)f_{2R}(r,T) \Big) \nonumber \\
& - 2m_{\Delta}r \tilde{\beta}_2 v(r,T) \Big( F^{\ast}_{1L}(r,T)f_{2R}(r,T) + F^{\ast}_{2L}(r,T)f_{1R}(r,T) \Big) \nonumber \\
& + 2r \beta_2 \Big( (\partial_r F^{\ast}_{1L}(r,T))f_{1L}(r,T) - (\partial_r F^{\ast}_{2L}(r,T))f_{2L}(r,T) \Big) \nonumber \\
& + 2r \tilde{\beta}_2 v(r,T) \Big( (\partial_r F^{\ast}_{1L}(r,T))f_{2L}(r,T) + (\partial_r F^{\ast}_{2L}(r,T))f_{1L}(r,T) \Big) \bigg\} \\[5ex]
C_4(Q^2,T) ={}& -m^2_N \int \frac{dr}{r} A(Q,r,T) \Big[ 
(\alpha_2+2\beta_2) \big( F^{\ast}_{1L}(r,T)f_{1R}(r,T) - F^{\ast}_{2L}(r,T)f_{2R}(r,T) \big) \nonumber \\
& + (\tilde{\alpha}_2+2\tilde{\beta}_2)v(r,T) \big( F^{\ast}_{1L}(r,T)f_{2R}(r,T) + F^{\ast}_{2L}(r,T)f_{1R}(r,T) \big) \Big] 
\end{align}

\begin{align}
C_5(Q^2,T) ={}& \int dr \frac{\partial_rA(Q,r,T)}{r^2} \bigg\{ 
\kappa \Big( F^{\ast}_{1L}(r,T)f_{1R}(r,T) + F^{\ast}_{2L}(r,T)f_{2R}(r,T) \Big) \nonumber \\
& + \tilde{\kappa}v(r,T) \Big( F^{\ast}_{1L}(r,T)f_{2R}(r,T) - F^{\ast}_{2L}(r,T)f_{1R}(r,T) \Big) \nonumber \\
& + r \alpha_2 \Big( (\partial_r F^{\ast}_{1L}(r,T))f_{1R}(r,T) - (\partial_r F^{\ast}_{2L}(r,T))f_{2R}(r,T) \Big) \nonumber \\
& + r \tilde{\alpha}_2 v(r,T) \Big( (\partial_r F^{\ast}_{1L}(r,T))f_{2R}(r,T) + (\partial_r F^{\ast}_{2L}(r,T))f_{1R}(r,T) \Big) \bigg\} \nonumber \\
& - \int \frac{dr}{r} \partial^2_rA(Q,r,T) \Big[ \alpha_3 \big( F^{\ast}_{1L}(r,T)f_{1R}(r,T) - F^{\ast}_{2L}(r,T)f_{2R}(r,T) \big) \nonumber \\
& + \tilde{\alpha}_3 v(r,T) \big( F^{\ast}_{1L}(r,T)f_{2R}(r,T) + F^{\ast}_{2L}(r,T)f_{1R}(r,T) \big) \Big] \nonumber \\
& + 2m_{\Delta} \int \frac{dr}{r} \partial_rA(Q,r,T) \Big[ \beta_2 \big( F^{\ast}_{1L}(r,T)f_{1L}(r,T) - F^{\ast}_{2L}(r,T)f_{2L}(r,T) \big) \nonumber \\
& + \tilde{\beta}_2 v(r,T) \big( F^{\ast}_{1L}(r,T)f_{2L}(r,T) + F^{\ast}_{2L}(r,T)f_{1L}(r,T) \big) \Big]  \\[5ex]
C_6(Q^2,T) ={}& m^2_N \int \frac{dr}{r} A(Q,r,T) \Big[ 
\alpha_3 \big( F^{\ast}_{1L}(r,T)f_{1R}(r,T) - F^{\ast}_{2L}(r,T)f_{2R}(r,T) \big) \nonumber \\
& + \tilde{\alpha}_3 v(r,T) \big( F^{\ast}_{1L}(r,T)f_{2R}(r,T) + F^{\ast}_{2L}(r,T)f_{1R}(r,T) \big) \Big] 
\end{align}
$\;$

Here, $F(r,T)$ and $f(r,T)$ denote the nucleon and $\Delta$ baryon thermal profile functions, respectively, which satisfy the following relations: $f_{2R}^{(n)} \to f_{1L}^{(n)}$, $f_{2L}^{(n)} \to -f_{1R}^{(n)}$, $F_{2R}^{(n)} \to F_{1L}^{(n)}$, and $F_{2L}^{(n)} \to -F_{1R}^{(n)}$. Their explicit forms are given by:

\begin{align}
F_{n}^{L,R}(r,T) &= \sqrt{\frac{2\Gamma (n+1)}{\Gamma (n+m_{\frac{1}{2}(L,R)}+1)}} K^{m_{\frac{1}{2}(L,R)}+1} r^{m_{\frac{1}{2}(L,R)}+\frac{1}{2}} e^{-\frac{K^{2}r^{2}}{2}} L_{n}^{m_{\frac{1}{2}(L,R)}}\left(K^{2}r^{2}\right), \label{ff1} \\[5ex]
f_{n}^{L,R}(r,T) &= \sqrt{\frac{2\Gamma (n+1)}{\Gamma (n+m_{\frac{3}{2}(L,R)}+1)}} K^{m_{\frac{3}{2}(L,R)}+1} r^{m_{\frac{3}{2}(L,R)}+\frac{1}{2}} e^{-\frac{K^{2}r^{2}}{2}} L_{n}^{m_{\frac{3}{2}(L,R)}}\left(K^{2}r^{2}\right). \label{ff2}
\end{align}

\noindent  which were obtained in \cite{mamedov2021temperature, nasibova2022meson} for the thermal soft-wall model.  
Here, $m_{L,R}=m\pm\frac{1}{2}$ and $m({\frac{1}{2},\frac{3}{2}})=N+L-\frac{3}{2}$, where $N$ is the number of quark flavors and $L$ is the  orbital angular momentum  \cite{gutsche2019baryons}.

\clearpage

\section{Numerical Results}
\label{nu}

To initialize the numerical analysis at finite temperature,  $k$ in the soft-wall model is fixed at $k = 0.402\text{ GeV}$ following Gutsche et al. ~\cite{gutsche2019baryons, gutsche2019mesons}, corresponding to the pion decay constant $F_\pi = 0.087\text{ GeV}$.

The interaction lagrangian corresponding to the nucleon-$\Delta$ transition  (Eq.~\ref{lag})  involves the parameters  $(\kappa, \tilde{\kappa}, \alpha_{2}, \alpha_{3},   \tilde{\alpha}_{2},  \tilde{\alpha}_{3},  \beta_2,   \tilde{\beta}_2)$.   These are free parameters, that are model dependent. For instance, in the hard-wall model, these parameters were fixed in order to match form factors at $Q^2 = 0$ \cite{ahn2009spin}.  For the soft-wall model, we shall proceed in a similar manner and fix these values to match the latest experimental data and lattice QCD results, wherever  available. 
By setting the corresponding form factor values at zero momentum transfer, the model then predicts the complete form factor profiles for all other values of $Q^2$, at all temperatures until the critical phase transition point. The nucleon mass, $\Delta$ baryon mass and chiral condensate are $m_N = 0.94\text{ GeV}$, $m_\Delta = 1.21\text{ GeV}$, and $\sigma = (0.363\text{ GeV})^3$, respectively. 

\subsection{Electromagnetic  Form Factors for the $N \to \Delta$ Transition at Finite Temperatures}

For computing the   electromagnetic (EM) form factors of the $N \to \Delta$ transition, we take the soft-wall model parameters to be as follows:  
\begin{equation}
\{\kappa, \tilde{\kappa}, \alpha_2, \tilde{\alpha}_2, \alpha_3, \tilde{\alpha}_3, \beta_2, \tilde{\beta}_2\} = \{-5, -10, 0, 1.3, 0, 0.17, 0, 0\}  
\end{equation}
With this choice of parameters, the Jones–Scadron magnetic dipole $G_{M_1}$, electric quadrupole $G_{E_2}$, and Coulomb quadrupole $G_{C_2}$ form factors assume the following values at zero momentum transfer ($Q^{2}=0$) and zero temperature:   
\begin{equation}
G_{M_1}(0,0) \approx 3.80, \quad G_{E_2}(0,0) \approx 0.074, \quad \text{and} \quad G_{C_2}(0,0) \approx 0.66.
\end{equation}
Our model parameters are chosen in a way to fit these initial values of the form factors. The latter are the ones which have been reported in the literature from accumulated  experimental measurements and lattice QCD simulations    \cite{alexandrou2007axial, ramalho2009electromagnetic, ramalho2009nucleon, segovia2014nucleon, nicmorus2010delta}.

\begin{figure}[htbp]
   
    \includegraphics[width=0.53\textwidth]{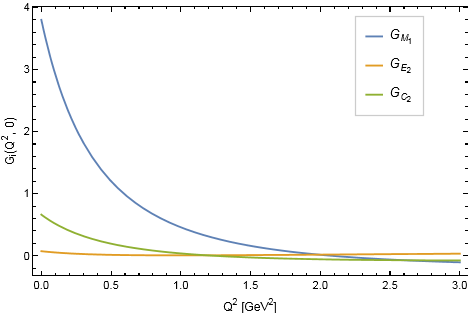} \hfill
    \includegraphics[width=0.45\textwidth]{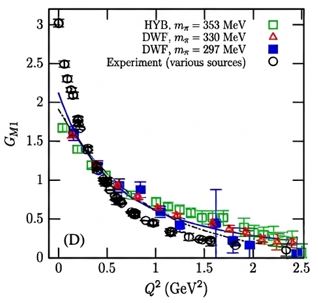} \par \hspace{2em} {$(a)$}  \hspace{23em}
    {$(b)$ }\par\vspace{0.5em}
    \includegraphics[width=0.45\textwidth]{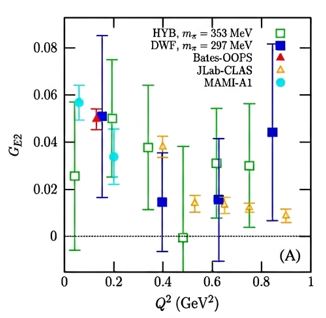} \hfill
    \includegraphics[width=0.45\textwidth]{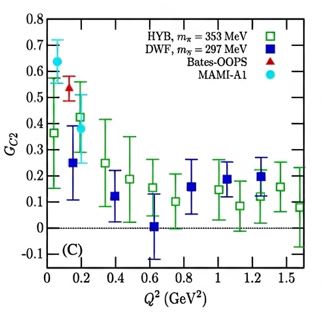} \par \hspace{2em} {$(c)$}  \hspace{23em}
    {$(d)$ }\par\vspace{1em}

\hspace{0.0\textwidth} 
\includegraphics[width=0.5\textwidth]{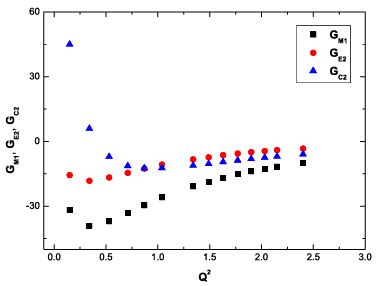}\par \hspace{1.5em}
    {$(e)$ }\par\vspace{0em}

\caption{$Q^{2}$ dependence of the form factors $G_{M_1}(Q^2, T)$, $G_{E_2}(Q^2, T)$ and $G_{C_2}(Q^2, T)$ at $T=0$ $\text{GeV}$. In panel (a) we show our thermal soft-wall model predictions; (b), (c) and (d) are lattice QCD data (DWF and hybrid action), alongside experimental results from Bates, JLab, and MAMI  (as indicated in the figures legends) \cite{alexandrou2005study, alexandrou2007axial, alexandrou2008nucleon, ramalho2009electromagnetic, mertz2001search, sparveris2003measurement, sparveris2007determination, sparveris2005investigation, joo2002q, a12006lowest}, and the hard-wall model  \cite{ahn2009spin} (e).}
    \label{1pp}
\end{figure}

Fig.~\ref{1pp}  illustrates the $Q^2$ dependence of the electromagnetic $N \to \Delta$ transition form factors and quadrupole ratios at zero temperature ($T = 0$). In panel~(a) we present the general $Q^2$ behavior of the transition form factors $G_{M_1}(Q^2, T)$, $G_{E_2}(Q^2, T)$, and $G_{C_2}(Q^2, T)$ evaluated within the  thermal soft-wall AdS/QCD framework. Panel~(b) displays the dominant magnetic dipole form factor $G_{M_1}(Q^2)$, where our theoretical model curve is comparable with experimental measurements compiled from Bates, JLab, and MAMI \cite{alexandrou2007axial}. Panel~(c) shows the subdominant electric quadrupole $G_{E_2}(Q^2)$, while panel~(d) depicts the Coulomb quadrupole $G_{C_2}(Q^2)$ form factors, all at zero temperature.  
The comparison data points in panels~(c) and (d) are extracted from coarse and fine lattice QCD simulations employing domain wall fermions and hybrid actions, alongside finite-temperature lattice calculations reported by Alexandrou et al. at $T = 0.1\text{ GeV}$ \cite{alexandrou2007axial}.  By appropriately tuning the soft-wall model parameters, our holographic predictions for the three form factors are brought into strong quantitative agreement with both, the extracted experimental measurements, and the lattice QCD benchmark data. Panel (e) illustrates the AdS/QCD hard-wall model  predictions for these form factors \cite{ahn2009spin}, demonstrating  trends somewhat comparable to our results.

Our results indicate that all electromagnetic form factors of the $N \to \Delta$ transition decrease monotonically with increasing momentum transfer. Furthermore, we see a considerably faster suppression of the quadrupole form factors $G_{E_2}$ and $G_{C_2}$. This indicates that the nucleon-to-$\Delta$ deformation is predominantly generated by long range pion-cloud dynamics. However, at large $Q^2$, the transition becomes increasingly governed by compact valence-quark degrees of freedom, in agreement with the asymptotic behavior predicted by perturbative QCD.

\FloatBarrier

\begin{figure}[htbp]
    \centering
   \includegraphics[scale=0.7]{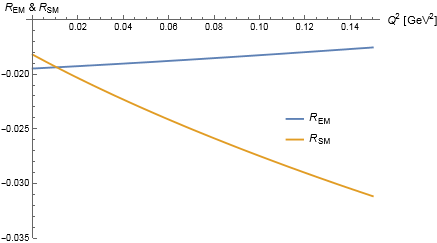}
   \includegraphics[scale=0.8]{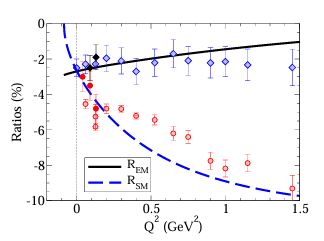} 
    \par \hspace{2em}   {$(a)$}\hspace{21em}
    {$(b)$ }\par\vspace{1em}
    \includegraphics[scale=0.5]{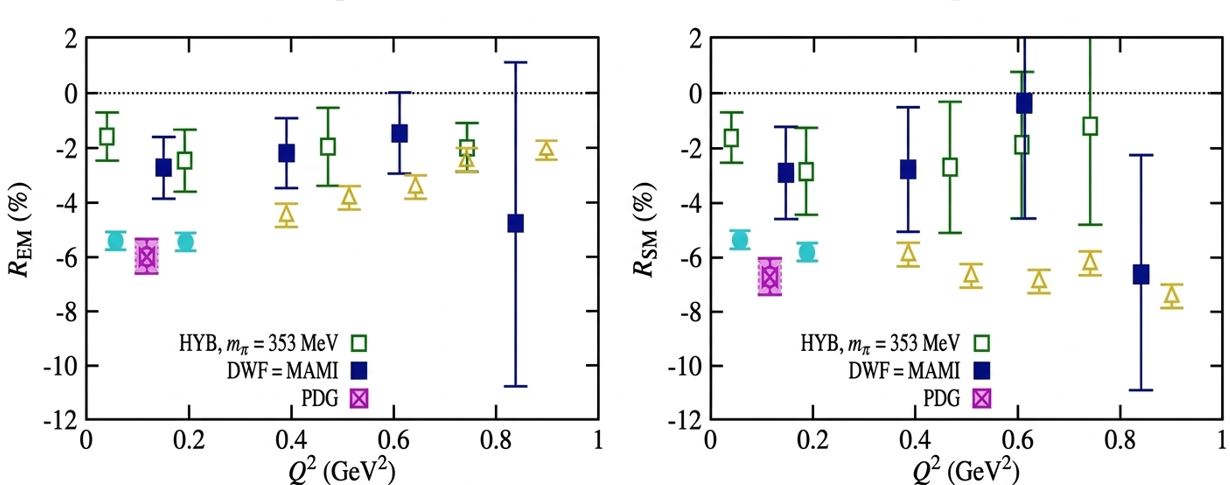} 
     \par \hspace{2em} {$(c)$}  \hspace{21em}
    {$(d)$}\par\vspace{1em}
      
\caption{$Q^{2}$ dependence of the $
R_{EM}$ and $R_{SM}$ ratios in the thermal soft-wall  at  ${T=0}$  $GeV$  (a),  and  lattice QCD /  experimental results (b), (c) and (d)  \cite{ramalho2009nucleon, alexandrou2007axial, alexandrou2008nucleon, ramalho2009electromagnetic, mertz2001search, sparveris2003measurement, sparveris2007determination, sparveris2005investigation, joo2002q, a12006lowest}.}
    \label{13a}
\end{figure}

Fig.~\ref{13a} shows the  electric $R_{\text{EM}}$ and Coulomb $R_{\text{SM}}$ quadrupole ratios. Panel~(a) shows our result within the soft-wall model, while panels~(b), (c) and (d) show the corresponding ratios estimated  from lattice QCD and experimental results \cite{ramalho2009nucleon, alexandrou2007axial, alexandrou2008nucleon, ramalho2009electromagnetic, mertz2001search, sparveris2003measurement, sparveris2007determination, sparveris2005investigation, joo2002q, a12006lowest}. Once again, we find good agreement with other QCD collaborations computing these ratios, thus validating the soft-wall model. 

\FloatBarrier

\begin{figure}[htbp]
    \centering
    \includegraphics[scale=0.6]{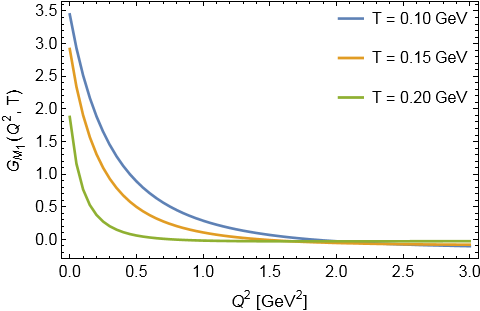} \hspace{1em}
    \includegraphics[scale=0.63]{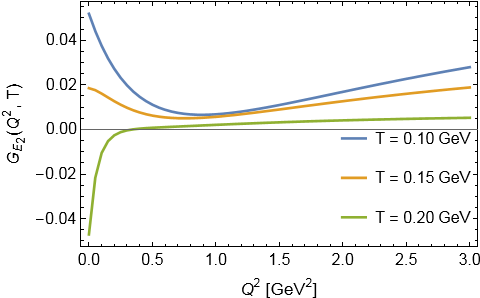}
    \par\textbf{$(a)$ $G_{M_1} (Q^2, T)$} \hspace{10em} \textbf{$(b)$ $G_{E_2} (Q^2, T)$}\par\vspace{2em}

    \includegraphics[scale=0.65]{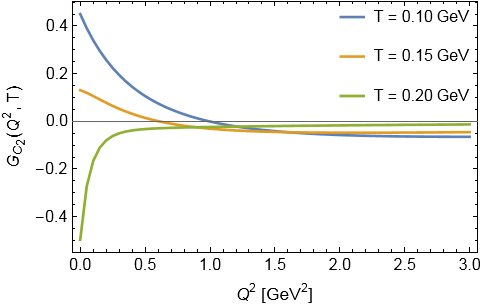}
    \par\textbf{$(c)$ $G_{C_2} (Q^2, T)$}\par\vspace{0.5em}

    \caption{ Momentum squared dependence of magnetic dipole, electric quadrupole and Coulomb
quadrupole form factors of the $N \to \Delta$ transition at different values of  $T$. }
.\label{2pp}
\end{figure}

Fig.~\ref{2pp} (a), (b) and (c) show our results for the $Q^2$ dependence of the $N \to \Delta$ transition form factors, namely,  $G_{M_1}(Q^2,T)$,  $G_{E_2}(Q^2,T)$, and  $G_{C_2}(Q^2,T)$, for three representative temperatures: $T = 0.10~\text{GeV}$, $T = 0.15~\text{GeV}$, and $T = 0.20~\text{GeV}$. Across all three panels, it is evident that for any given temperature, the magnitudes of the form factors decrease monotonically with increasing momentum transfer $Q^2$, while flattening out at higher values of $Q^2$. Regarding the thermal response, our numerical results demonstrate a clear thermal suppression effect across the entire $Q^2$ range: as the temperature increases from $T = 0.10~\text{GeV}$ to $T = 0.20~\text{GeV}$, the absolute values $|G_{M_1}(Q^2,T)|$, $|G_{E_2}(Q^2,T)|$, and $|G_{C_2}(Q^2,T)|$ systematically drop toward zero. Consequently, the curves corresponding to higher temperatures lie closer to the zero axis. Within our soft-wall AdS/QCD model, this suppression behavior originates from the temperature-dependent modification of the thermal background geometry and the dilaton profile. As the temperature rises toward the phase transition region, thermal fluctuations attenuate the effective bound-state wave function overlap between the nucleon and the $\Delta$ baryon in the 5D bulk, leading to a weaker transition amplitude. Furthermore, a comparison among the form factors shows that the dominant spin-flip channel $G_{M_1}(Q^2,T)$ retains the largest numerical scale, whereas the non-spherical quadrupole deformation channels $G_{E_2}(Q^2,T)$ and $G_{C_2}(Q^2,T)$ are smaller in magnitude and exhibit a rapid fall-off at short distances. 

\FloatBarrier

\begin{figure}[htbp]
    \centering
    \includegraphics[scale=0.6]{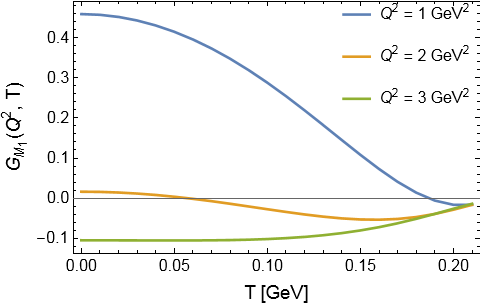} \hspace{1em}
    \includegraphics[scale=0.62]{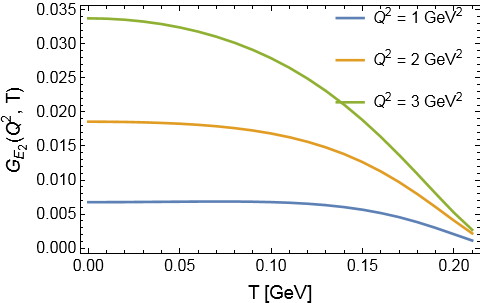}
    \par\textbf{$(a)$ $G_{M_1} (Q^2, T)$} \hspace{10em} \textbf{$(b)$ $G_{E_2} (Q^2, T)$}\par\vspace{2em}

    \includegraphics[scale=0.65]{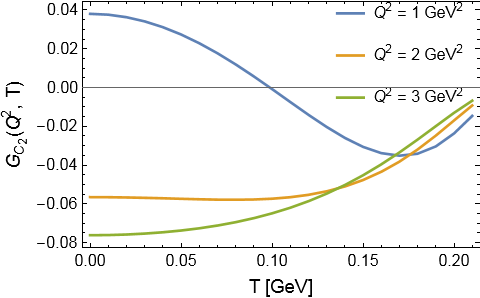}
    \par\textbf{$(c)$ $G_{C_2} (Q^2, T)$}\par\vspace{0.5em}

    \caption{Temperature  dependence of magnetic dipole, electric quadrupole and Coulomb
quadrupole form factors of the $N \to \Delta$  transition at different value of $Q^{2}$. }
      \label{3pp}
\end{figure}

Figure~\ref{3pp} displays the explicit temperature dependence of the $N \to \Delta$ transition form factors $G_{M_1}(Q^2,T)$, $G_{E_2}(Q^2,T)$, and $G_{C_2}(Q^2, T)$ at fixed values of the momentum transfer, specifically $Q^2 = 1~\text{GeV}^2$, $Q^2 = 2~\text{GeV}^2$, and $Q^2 = 3~\text{GeV}^2$.

Across all panels, a universal feature is the thermal suppression of the form factors as the temperature approaches the critical deconfinement regime ($T \to T_c \approx 0.22~\text{GeV}$). At low temperatures ($T \approx 0$), the form factors maintain their characteristic vacuum magnitudes, with the largest absolute strength observed at $Q^2 = 1~\text{GeV}^2$ (blue curves). As $Q^2$ increases to $2~\text{GeV}^2$ (orange  curves) and $3~\text{GeV}^2$ (green curves), the overall magnitudes are  suppressed due to short-distance power-law behavior.  

As the temperature increases and the square of the momentum transfer $Q^2$ grows from zero, the magnitudes of these form factors experience a monotonic suppression. Near the critical deconfinement temperature $T_c$, the form factors undergo a rapid variation and approach critical behavior at $T_c$. This temperature-induced suppression and critical behavior directly reflect the dehadronization process, indicating a phase transition from hadronic matter to the quark-gluon plasma (QGP).   
From the AdS bulk perspective, with increasing temperature, thermal fluctuations gradually weaken the bound-state spatial overlap in the 5D bulk geometry.  Consequently, regardless of the momentum transfer scale $Q^2$, all three form factors, $G_{M_1}(Q^2,T)$, $G_{E_2}(Q^2,T)$ and $G_{C_2}(Q^2,T)$ smoothly approach zero near $T_c \approx 0.22~\text{GeV}$.

\FloatBarrier

\begin{figure}[htbp]
    \centering
    \includegraphics[scale=0.55]{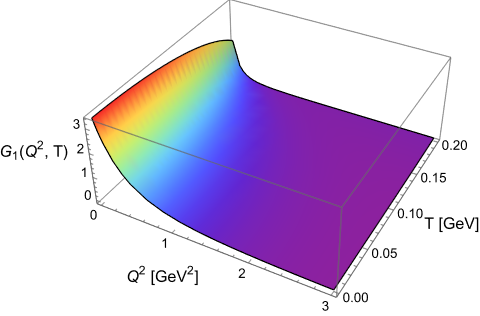} \hspace{1em}
    \includegraphics[scale=0.55]{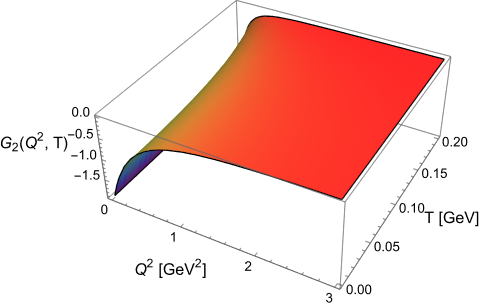}
    \par\textbf{$(a)$ $G_{1} (Q^2, T)$} \hspace{10em} \textbf{$(b)$ $G_{2} (Q^2, T)$}\par\vspace{2em}
    \includegraphics[scale=0.55]{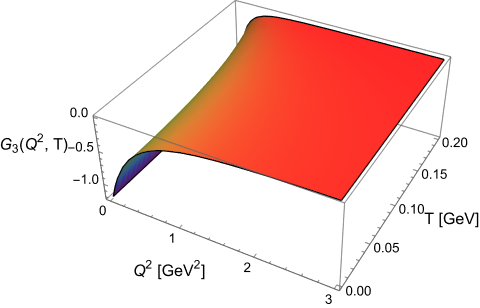}
    \par\textbf{$(c)$ $G_{3} (Q^2, T)$}\par\vspace{0.3em}

 \caption{Temperature and momentum squared dependence of the invariant form factors \( G_{{1}}(Q^2, T) \),   \( G_{{2}}(Q^2, T) \), and   \( G_{{2}}(Q^2, T) \) of the $N \to \Delta$ transition.}
      \label{7a}
\end{figure}

\begin{figure}[htbp]
    \centering
    \includegraphics[scale=0.55]{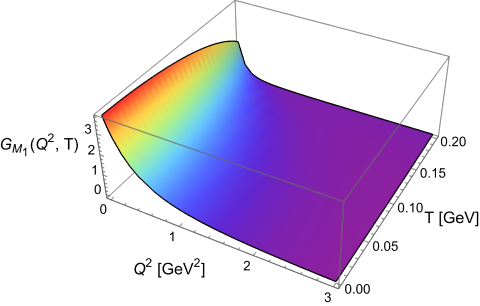} \hspace{1em}
    \includegraphics[scale=0.55]{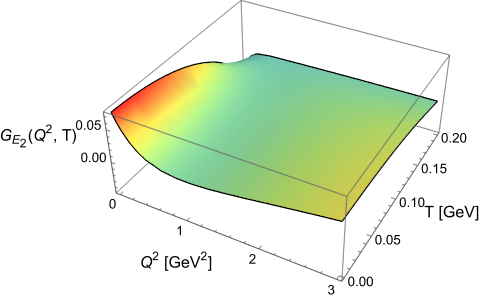}
    \par\textbf{$(a)$ $G_{M_1} (Q^2, T)$} \hspace{10em} \textbf{$(b)$ $G_{E_2} (Q^2, T)$}\par\vspace{2em}
    \includegraphics[scale=0.55]{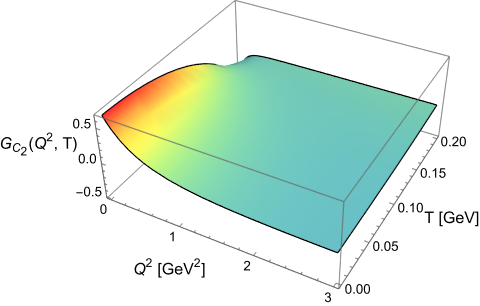}
    \par\textbf{$(c)$ $G_{C_2} (Q^2, T)$}\par\vspace{0.5em}
 \caption{Temperature and momentum squared dependence of  \( G_{M_{1}}(Q^2, T) \),   \( G_{E_{2}}(Q^2, T) \)  and   \( G_{C_{2}}(Q^2, T) \)  form factors of the $N \to \Delta$ transition. }
      \label{2p}
\end{figure}

In Fig.~\ref{7a} and Fig.~\ref{2p}, we present $T$ and $Q^{2}$ dependent $3D$ graphs of the   vector form factors:    \( G_{{1}}(Q^2, T) \),   \( G_{{2}}(Q^2, T) \),   \( G_{{2}}(Q^2, T) \) and  \( G_{M_{1}}(Q^2, T) \),   \( G_{E_{2}}(Q^2, T) \),   \( G_{C_{2}}(Q^2, T) \)  respectively, for the $N \to \Delta$ transition. These plots show that with increasing values of $T$ and $Q^{2}$, the values of all the form factors systematically decrease.

\FloatBarrier

\FloatBarrier
Table~\ref{tab1} shows the electromagnetic (vector) form factor values for various temperature and momentum squared, chosen in accordance with the selection in the hard-wall model \cite{ahn2009spin}, for comparison with the latter.

\begin{table}[htbp]
\centering
\caption{The values of $Q^2$ vs $T$ dependence of $G_{M_{1}}, G_{E_{2}}, G_{C_{2}}$ form-factors.}
\label{tab:form_factors}
\small
\setlength{\tabcolsep}{6pt}
\begin{tabular}{ccccc c ccccc}
\toprule
$Q^2$ & $T$ & $G_{M_{1}}$ & $G_{E_{2}}$ & $G_{C_{2}}$ & & $Q^2$ & $T$ & $G_{M_{1}}$ & $G_{E_{2}}$ & $G_{C_{2}}$ \\
\cmidrule{1-5} \cmidrule{7-11}
0.00 & 0.0 & 3.8009 & 0.0741 & 0.6630 & & 1.49 & 0.0 & 0.1639 & 0.0110 & -0.0261 \\
0.00 & 0.1 & 3.4478 & 0.0518 & 0.4500 & & 1.49 & 0.1 & 0.0722 & 0.0108 & -0.0407 \\
0.00 & 0.2 & 1.8775 & -0.0471 & -0.4967 & & 1.49 & 0.2 & -0.0277 & 0.0032 & -0.0199 \\
\addlinespace
0.15 & 0.0 & 2.5816 & 0.0453 & 0.4563 & & 1.63 & 0.0 & 0.1127 & 0.0129 & -0.0370 \\
0.15 & 0.1 & 2.1724 & 0.0319 & 0.2959 & & 1.63 & 0.1 & 0.0368 & 0.0124 & -0.0469 \\
0.15 & 0.2 & 0.5297 & -0.0053 & -0.1108 & & 1.63 & 0.2 & -0.0285 & 0.0034 & -0.0190 \\
\addlinespace
0.34 & 0.0 & 1.6750 & 0.0246 & 0.2884 & & 1.77 & 0.0 & 0.0706 & 0.0150 & -0.0457 \\
0.34 & 0.1 & 1.3095 & 0.0175 & 0.1730 & & 1.77 & 0.1 & 0.0083 & 0.0141 & -0.0517 \\
0.34 & 0.2 & 0.1599 & -0.0000 & -0.0441 & & 1.77 & 0.2 & -0.0289 & 0.0037 & -0.0182 \\
\addlinespace
0.53 & 0.0 & 1.1311 & 0.0137 & 0.1803 & & 1.90 & 0.0 & 0.0378 & 0.0170 & -0.0523 \\
0.53 & 0.1 & 0.8307 & 0.0103 & 0.0959 & & 1.90 & 0.1 & -0.0135 & 0.0156 & -0.0552 \\
0.53 & 0.2 & 0.0504 & 0.0008 & -0.0313 & & 1.90 & 0.2 & -0.0289 & 0.0039 & -0.0175 \\
\addlinespace
0.71 & 0.0 & 0.7965 & 0.0088 & 0.1106 & & 2.03 & 0.0 & 0.0101 & 0.0190 & -0.0577 \\
0.71 & 0.1 & 0.5528 & 0.0073 & 0.0476 & & 2.03 & 0.1 & -0.0316 & 0.0172 & -0.0579 \\
0.71 & 0.2 & 0.0092 & 0.0013 & -0.0271 & & 2.03 & 0.2 & -0.0289 & 0.0041 & -0.0169 \\
\addlinespace
0.87 & 0.0 & 0.5877 & 0.0070 & 0.0659 & & 2.15 & 0.0 & -0.0119 & 0.0209 & -0.0618 \\
0.87 & 0.1 & 0.3868 & 0.0066 & 0.0175 & & 2.15 & 0.1 & -0.0457 & 0.0186 & -0.0599 \\
0.87 & 0.2 & -0.0085 & 0.0017 & -0.0250 & & 2.15 & 0.2 & -0.0287 & 0.0043 & -0.0163 \\
\addlinespace
1.04 & 0.0 & 0.4251 & 0.0068 & 0.0306 & & 2.40 & 0.0 & -0.0489 & 0.0249 & -0.0683 \\
1.04 & 0.1 & 0.2621 & 0.0069 & -0.0056 & & 2.40 & 0.1 & -0.0688 & 0.0215 & -0.0627 \\
1.04 & 0.2 & -0.0184 & 0.0021 & -0.0233 & & 2.40 & 0.2 & -0.0281 & 0.0046 & -0.0152 \\
\addlinespace
1.34 & 0.0 & 0.2315 & 0.0091 & -0.0116 & & & & & & \\
1.34 & 0.1 & 0.1199 & 0.0092 & -0.0320 & & & & & & \\
1.34 & 0.2 & -0.0261 & 0.0028 & -0.0210 & & & & & & \\
\bottomrule
\end{tabular}
\label{tab1}
\end{table}

\FloatBarrier

\FloatBarrier

\subsection{Axial Vector Form Factors for the $N \to \Delta$ Transition at Finite Temperatures}

\begin{figure}[htbp]
    \centering
   \includegraphics[scale=0.5]{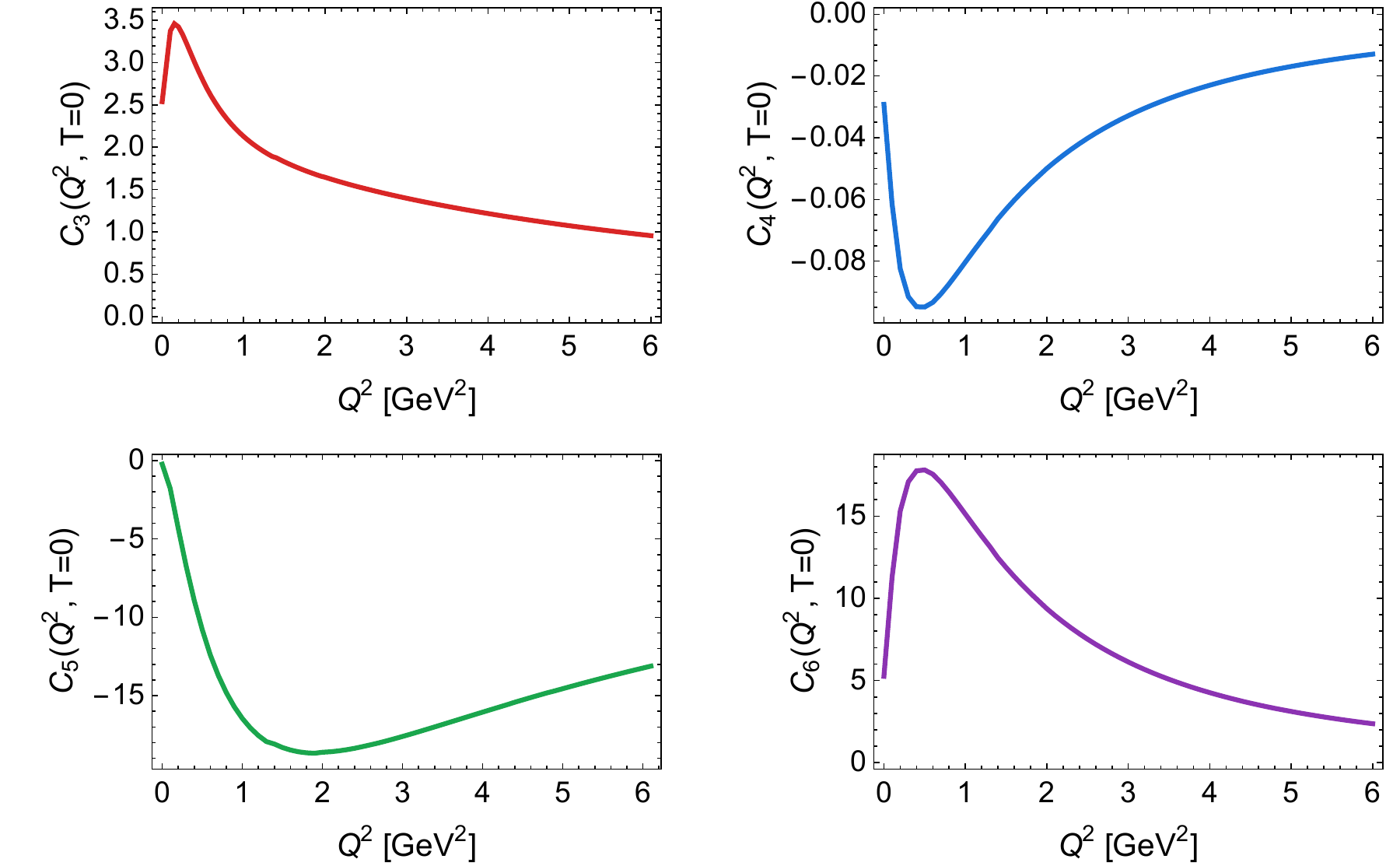}
     \par{$(a)$  } ${T=0}$  $GeV$  \\ 
\vspace{3em}  
   \includegraphics[scale=0.5]{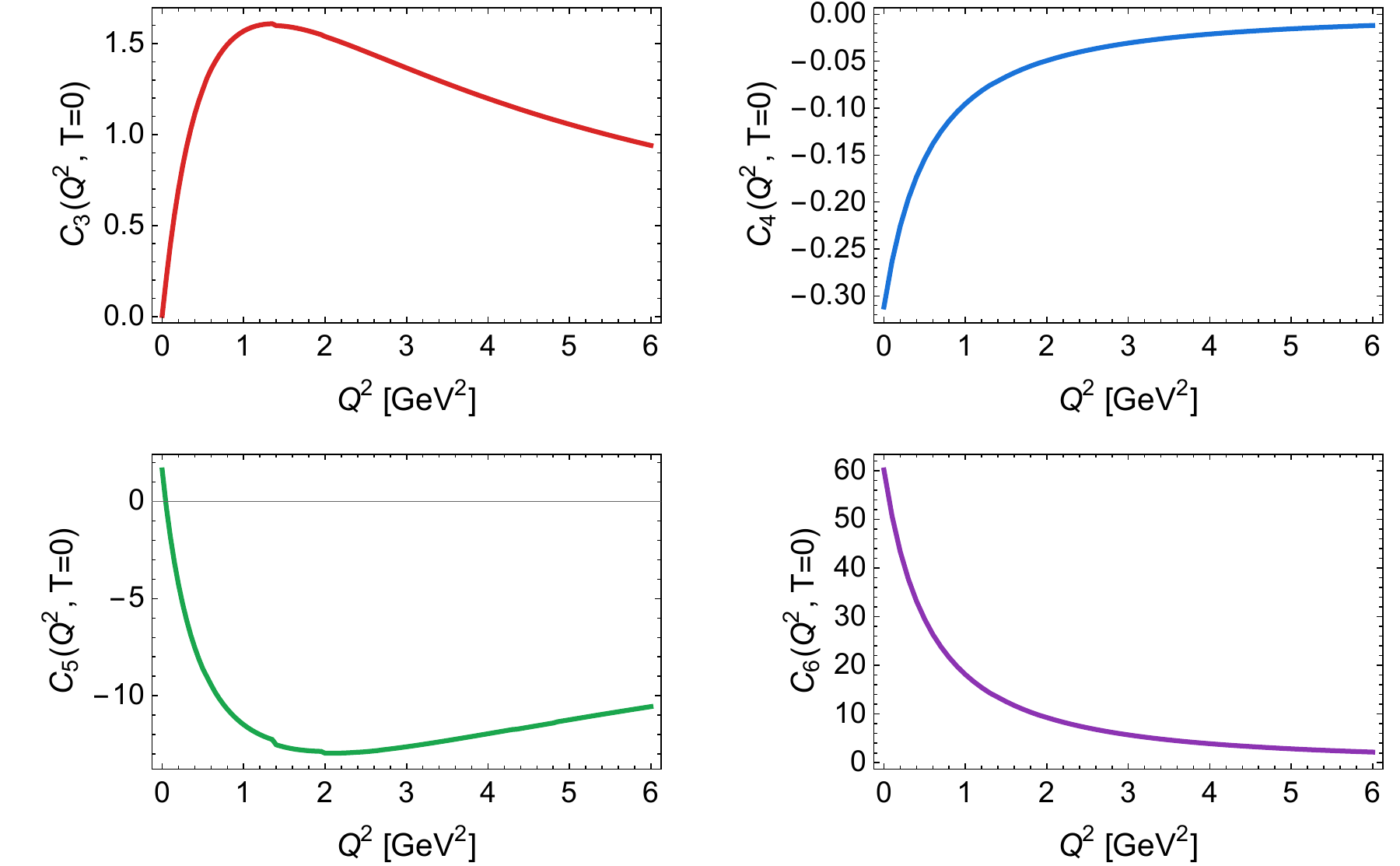}
     \par{$(b)$  } ${T=0}$  $GeV$ 
  \caption{$Q^2$ dependence of the axial vector form factors $C_i(Q^2, T=0)$ ($i=3,4,5,6$) for the $N \to \Delta$ transition. Results from the thermal soft-wall model using (a) the standard axial vector propagator, and (b) the parametrically deformed axial propagator.
  }   
    \label{7p}
\end{figure}

\begin{figure}[p] 
    \centering
    \begin{minipage}{\textwidth}
        \centering
        \includegraphics[width=0.9\textwidth]{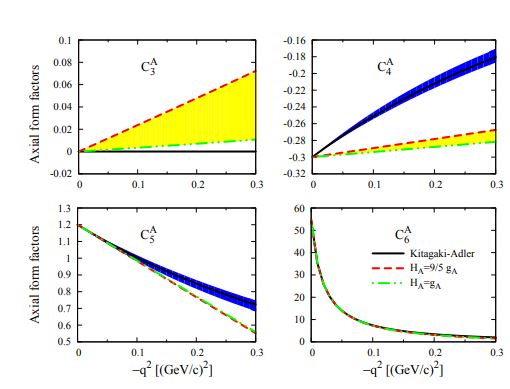}\\[0.3em]
        (a) $T = 0\text{ GeV}$
        \vspace{3em} \\ 
       \includegraphics[width=0.4\textwidth]{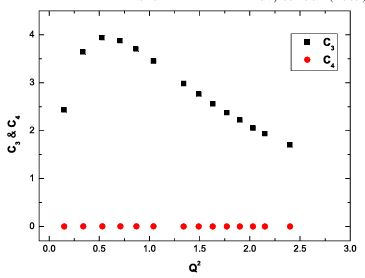}\hspace{1em}%
        \includegraphics[width=0.4\textwidth]{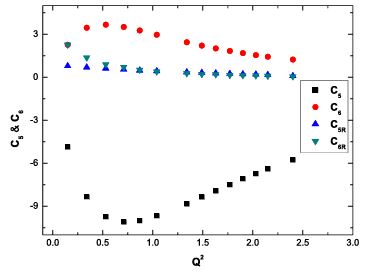}\\[0.3em]
        (b) $T = 0\text{ GeV}$
        \vspace{1em}
\caption{Literature results showing the $Q^2$ profiles of axial form factors $C_i(Q^2, T=0)$ ($i=3,4,5,6$) for $N \to \Delta$: (a) from relativistic chiral perturbation theory using the Kitagaki--Adler parametrization \cite{geng2008nucleon}; (b) from the hard-wall AdS/QCD model \cite{ahn2009spin}, where black squares and red circles represent $C_5^A(Q^2)$ and $C_6^A(Q^2)$, respectively, while blue triangles and green inverted triangles denote the lattice QCD results from \cite{alexandrou2007axial}. } 
\label{7plit}  
    \end{minipage} 
\end{figure}

\begin{figure}[htbp]
    \centering
    \begin{minipage}{0.48\textwidth}
        \centering
        \includegraphics[width=\textwidth]{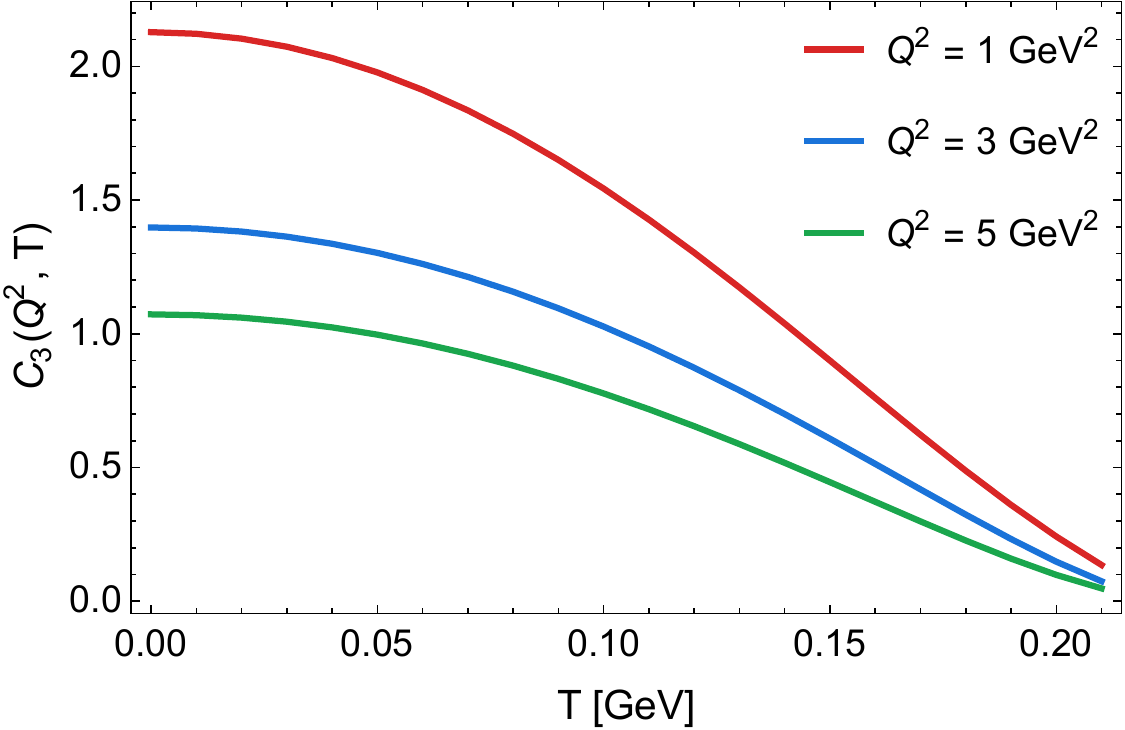} \\
        (a) $C_3(Q^2, T)$
    \end{minipage}
    \hfill
    \begin{minipage}{0.48\textwidth}
        \centering
        \includegraphics[width=\textwidth]{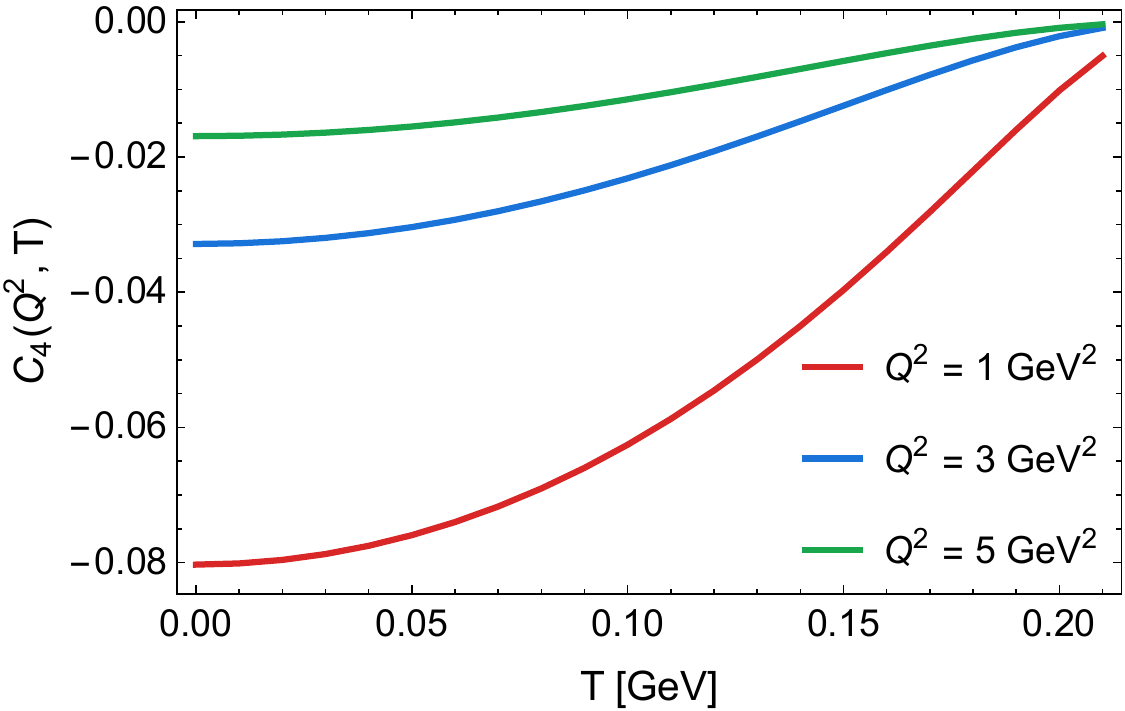} \\
        (b) $C_4(Q^2, T)$
    \end{minipage}

    \vspace{1em}

    \begin{minipage}{0.48\textwidth}
        \centering
        \includegraphics[width=\textwidth]{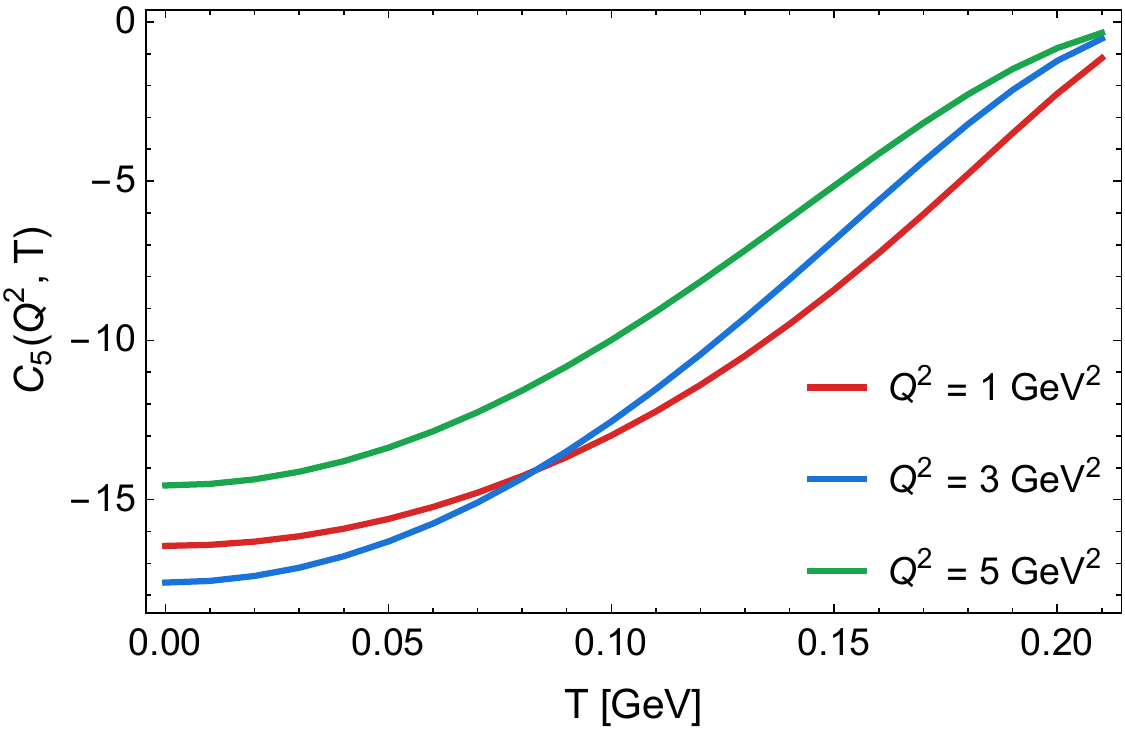} \\
        (c) $C_5(Q^2, T)$
    \end{minipage}
    \hfill
    \begin{minipage}{0.48\textwidth}
        \centering
        \includegraphics[width=\textwidth]{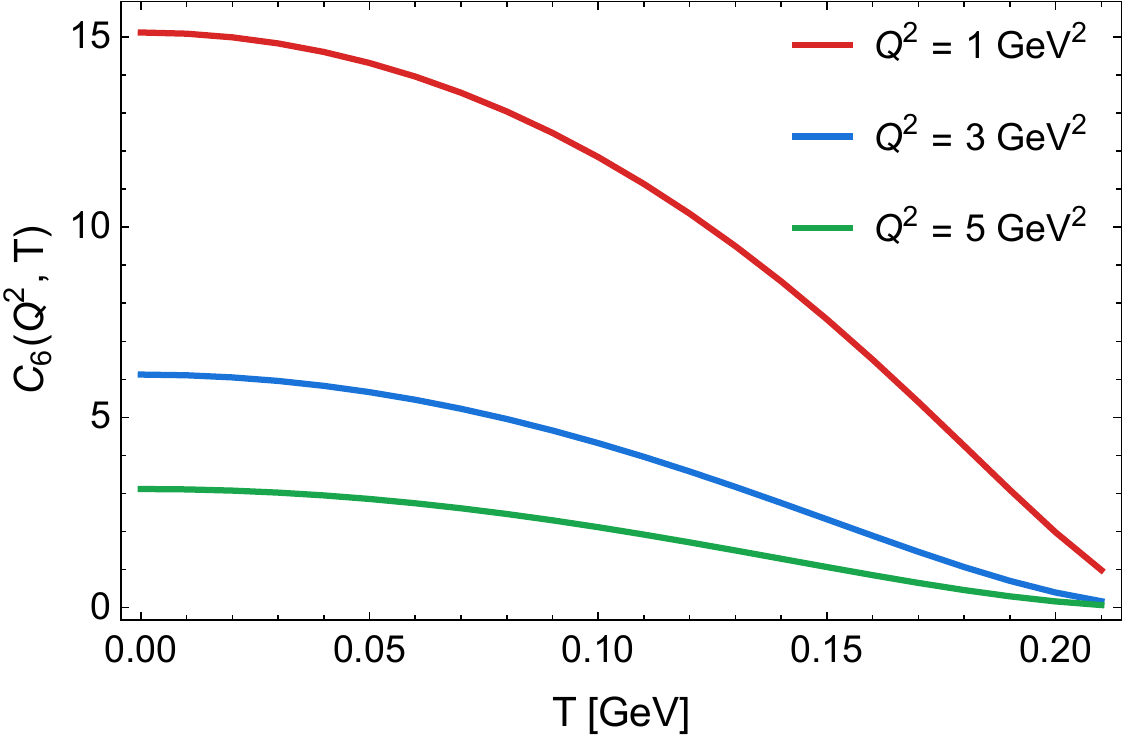} \\
        (d) $C_6(Q^2, T)$
    \end{minipage}

    \caption{$T$ dependence of the axial vector form factors $C_3(Q^2,T)$, $C_4(Q^2,T)$, $C_5(Q^2,T)$, and $C_6(Q^2,T)$ of the $N \to \Delta$ transition, based on the standard axial propagator, at different values of $Q^2$.}  
    \label{8pa}
\end{figure}

\begin{figure}[htbp]
    \centering
    \begin{minipage}{0.48\textwidth}
        \centering
        \includegraphics[width=\textwidth]{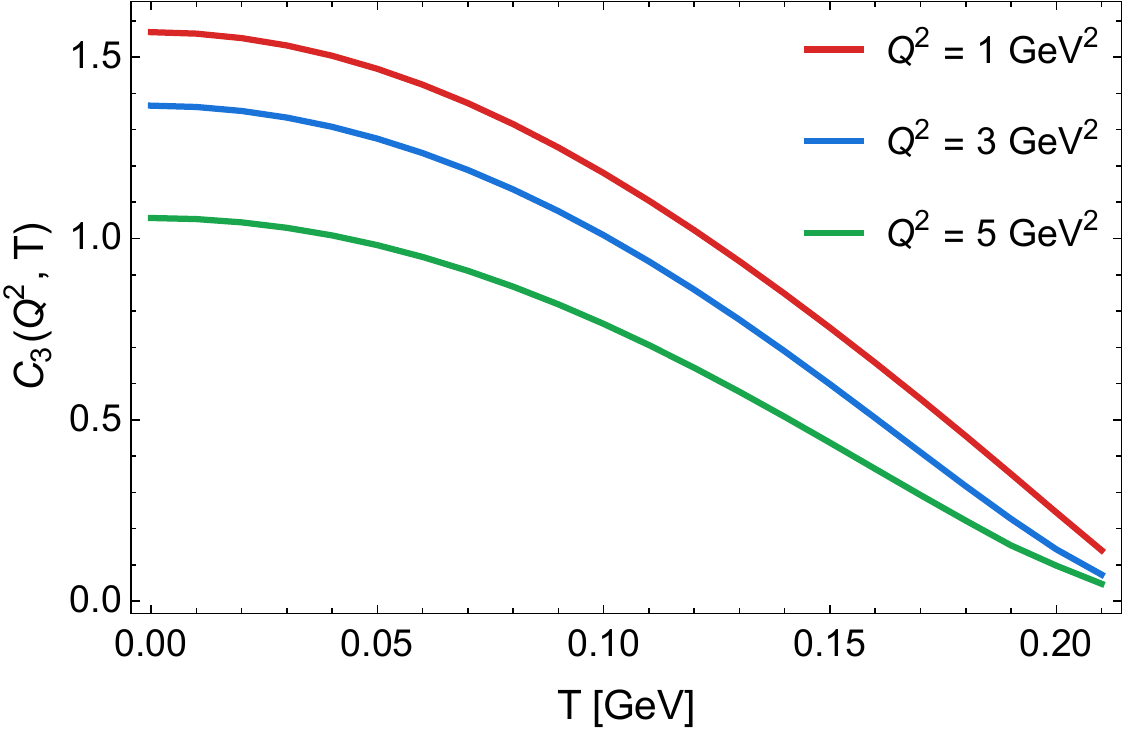} \\
        (a) $C_3(Q^2, T)$
    \end{minipage}
    \hfill
    \begin{minipage}{0.48\textwidth}
        \centering
        \includegraphics[width=\textwidth]{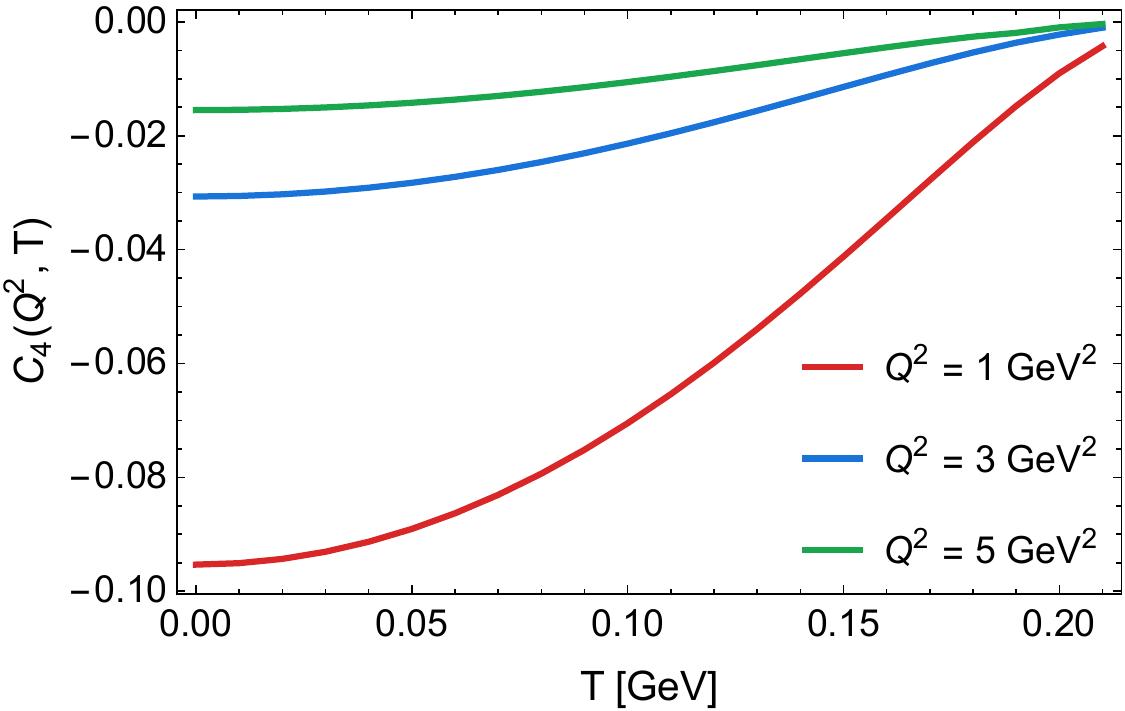} \\
        (b) $C_4(Q^2, T)$
    \end{minipage}

    \vspace{1em}

    \begin{minipage}{0.48\textwidth}
        \centering
        \includegraphics[width=\textwidth]{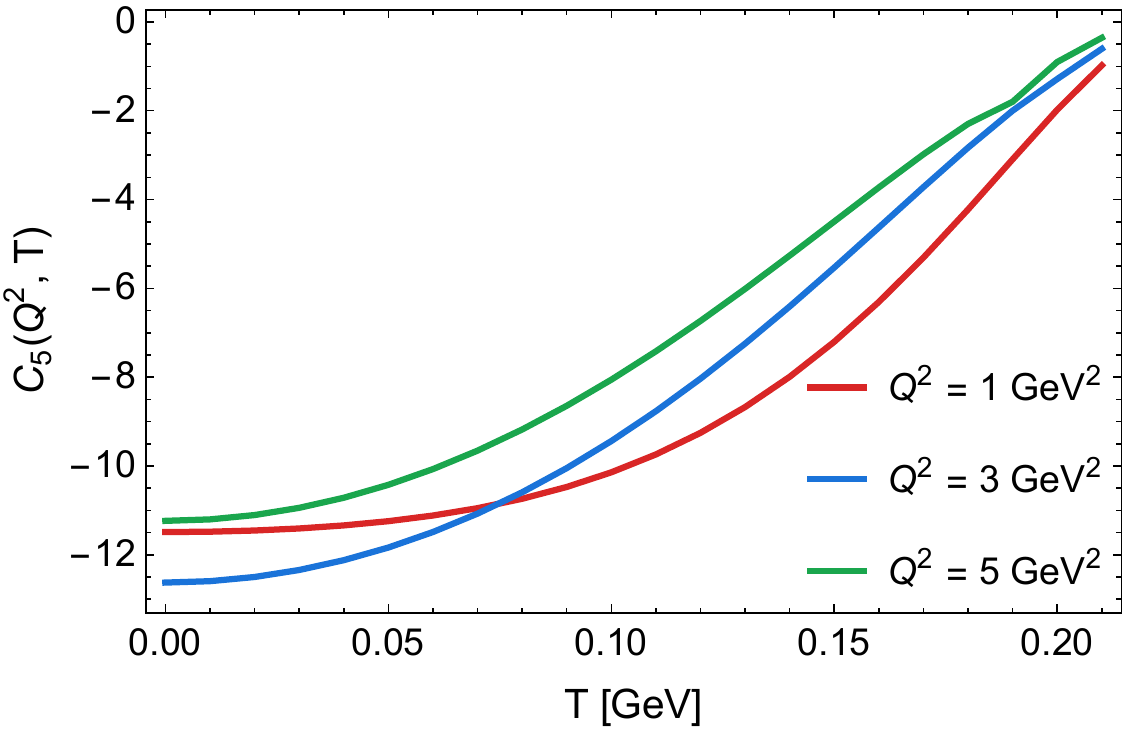} \\
        (c) $C_5(Q^2, T)$
    \end{minipage}
    \hfill
    \begin{minipage}{0.48\textwidth}
        \centering
        \includegraphics[width=\textwidth]{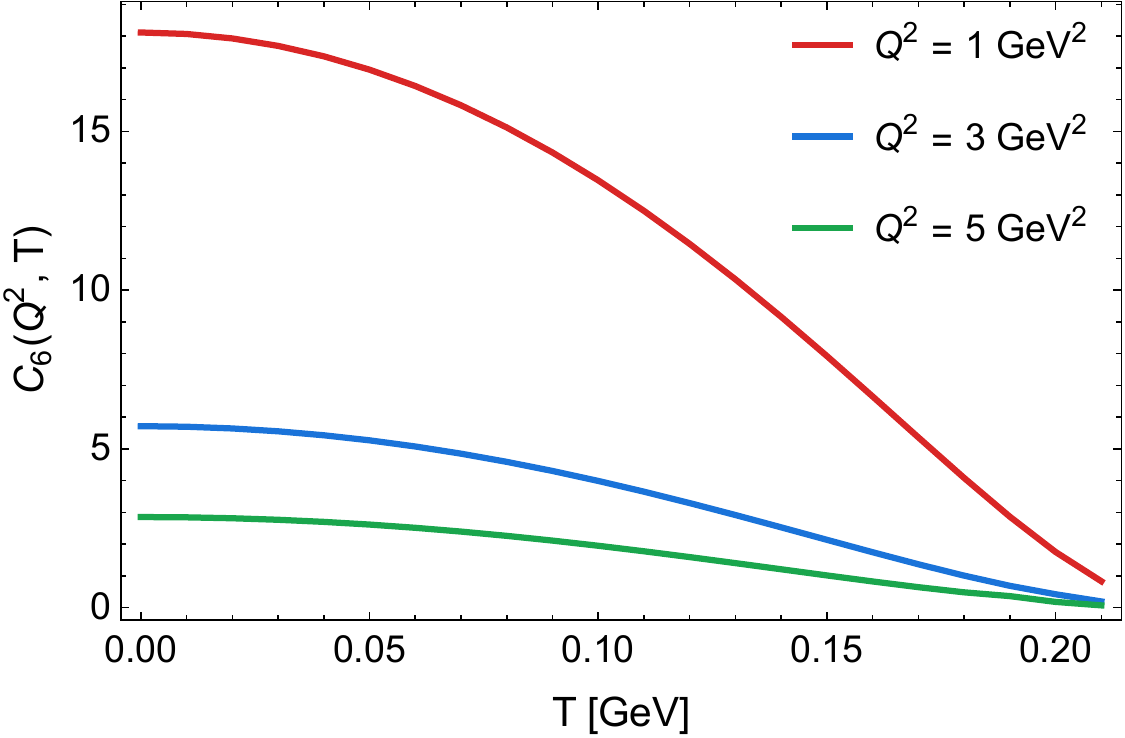} \\
        (d) $C_6(Q^2, T)$
    \end{minipage}

    \caption{$T$ dependence of the axial vector form factors $C_3(Q^2,T)$, $C_4(Q^2,T)$, $C_5(Q^2,T)$, and $C_6(Q^2,T)$ of the $N \to \Delta$ transition, based on the parametrically deformed axial propagator, at different values of $Q^2$.}
    \label{8p}
\end{figure}

\begin{figure}[htbp]
    \centering
    \begin{minipage}{0.48\textwidth}
        \centering
        \includegraphics[width=\textwidth]{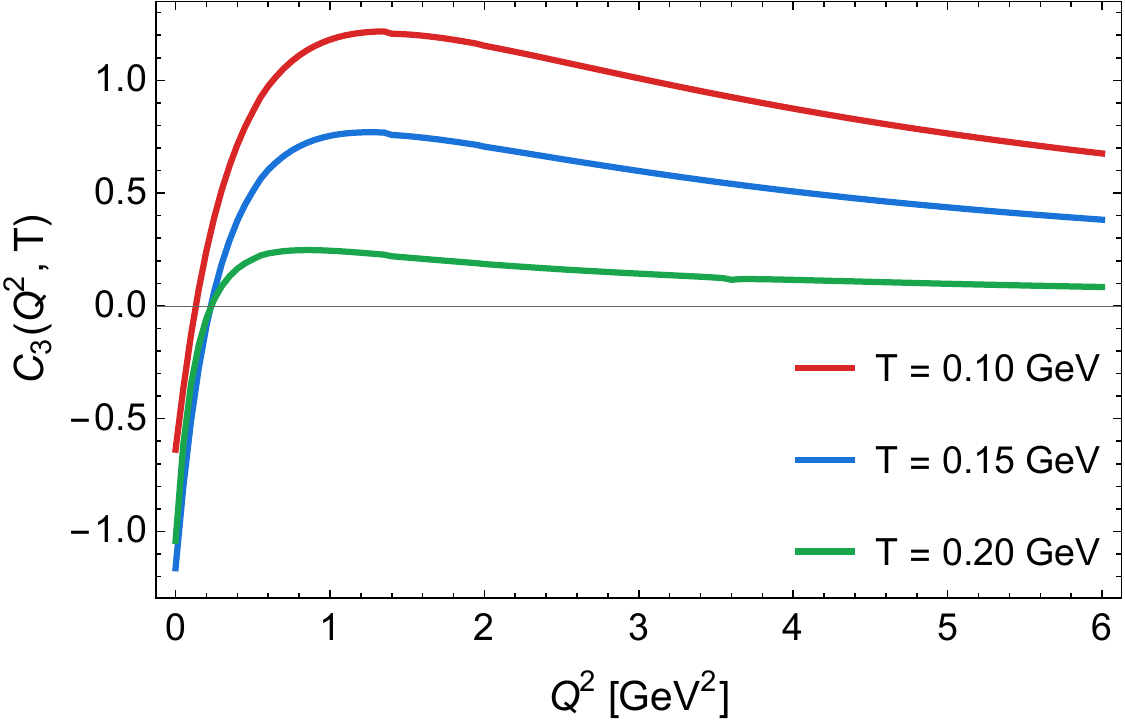} \\
        (a) $C_3(Q^2, T)$
    \end{minipage}
    \hfill
    \begin{minipage}{0.48\textwidth}
        \centering
        \includegraphics[width=\textwidth]{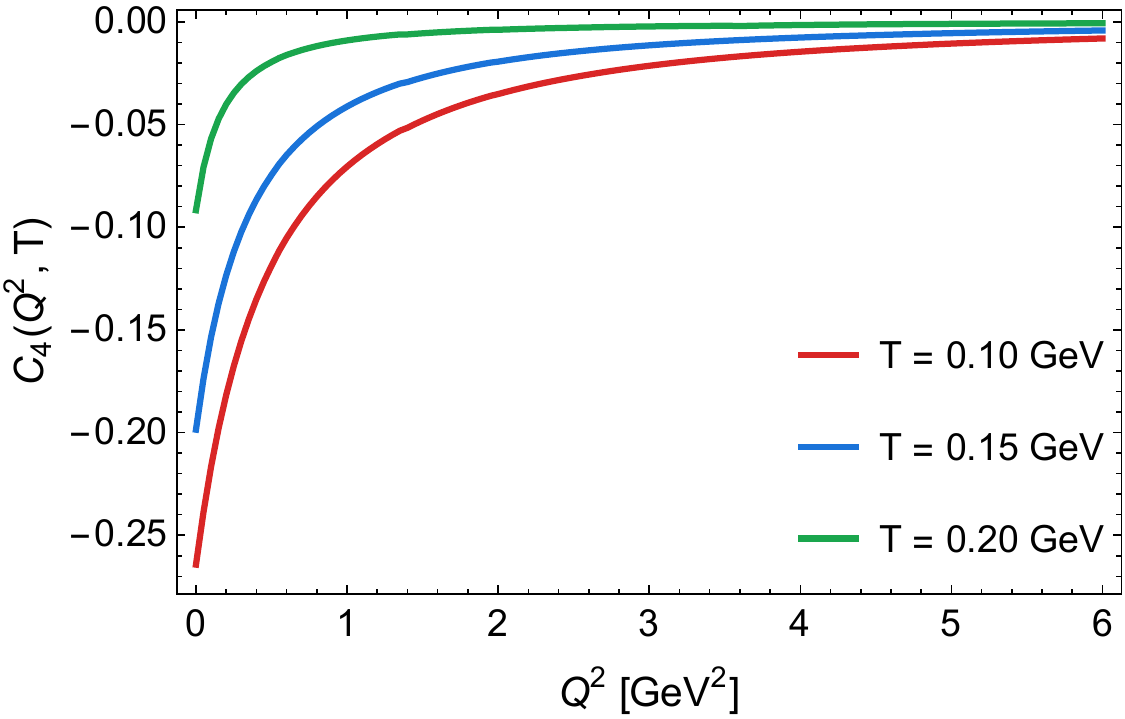} \\
        (b) $C_4(Q^2, T)$
    \end{minipage}

    \vspace{1em}

    \begin{minipage}{0.48\textwidth}
        \centering
        \includegraphics[width=\textwidth]{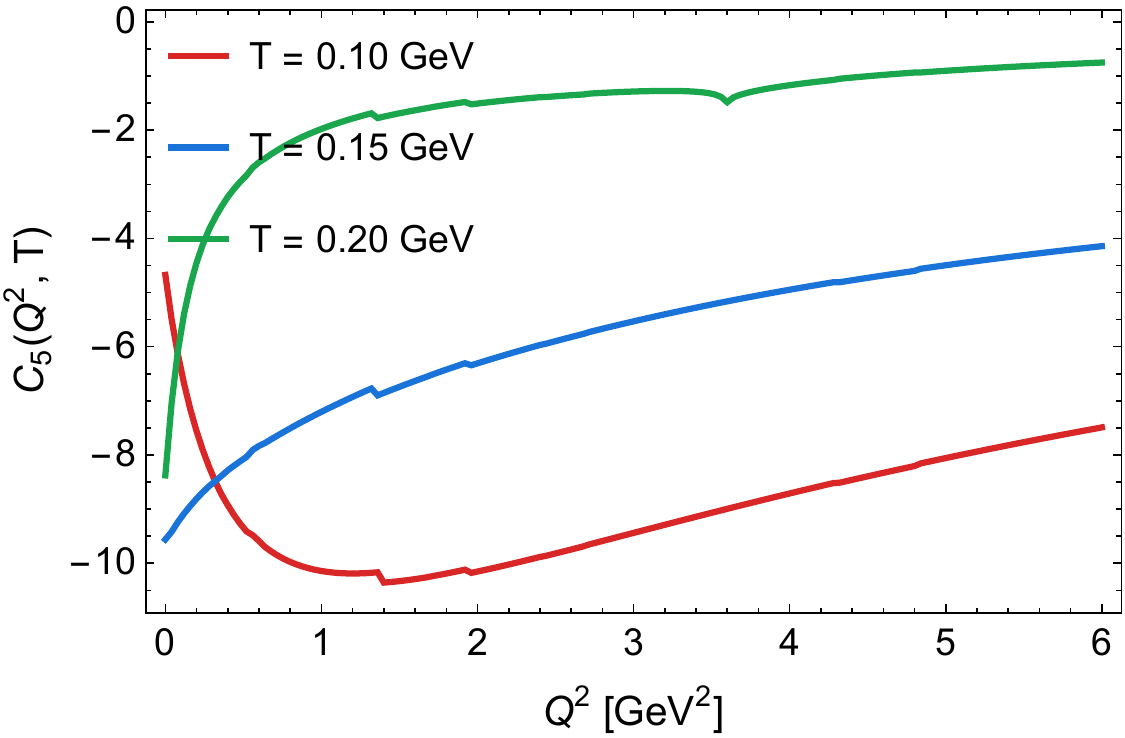} \\
        (c) $C_5(Q^2, T)$
    \end{minipage}
    \hfill
    \begin{minipage}{0.48\textwidth}
        \centering
        \includegraphics[width=\textwidth]{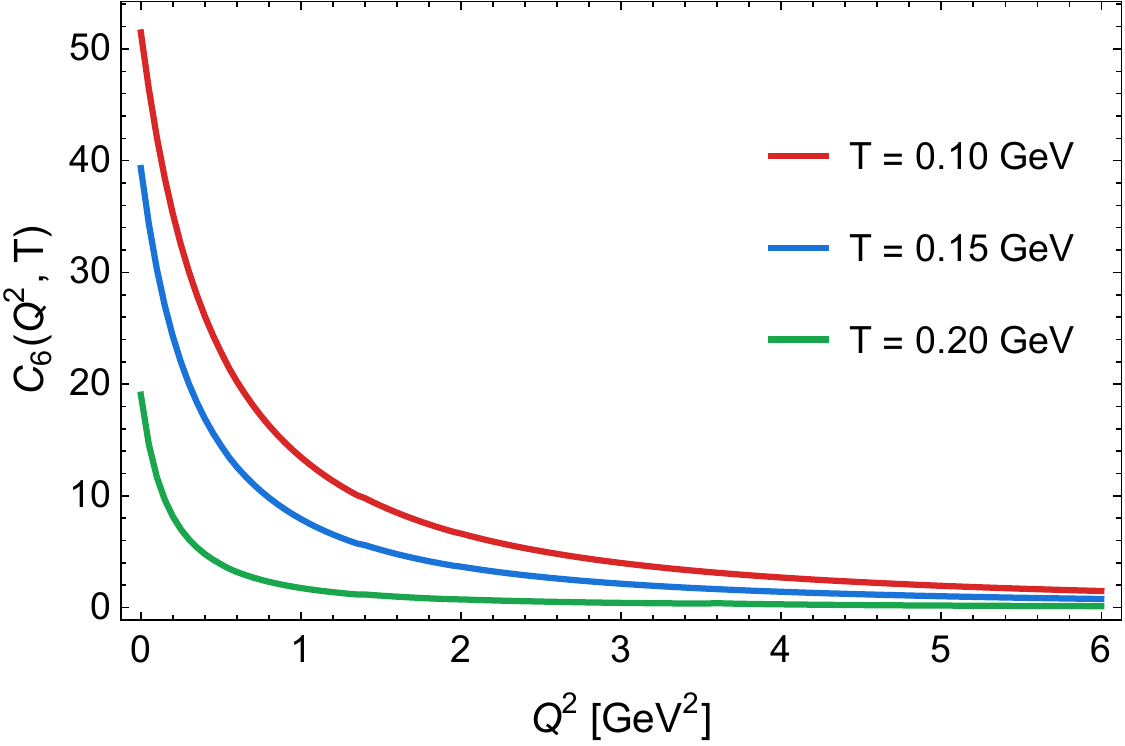} \\
        (d) $C_6(Q^2, T)$
    \end{minipage}

\caption{$Q^{2}$ dependence of the axial  vector form factors of the $N \to \Delta$  transition at different values of $T$  
based on the deformed axial propagator: (a) $C_3(Q^2,T)$, (b) $C_4(Q^2,T)$, (c) $C_5(Q^2,T)$, and (d) $C_6(Q^2,T)$.}
    \label{9p}
\end{figure}

The axial vector form factors for the $N \to \Delta$  transition characterize their spin-isospin structure, providing insight into the distribution of axial charge within the baryon. As the temperature increases, the axial vector form factors gradually decrease due to  thermal effects and reduction of wave function overlap, and they approach zero near the critical temperature, signaling the effective loss of axial charge and the dissolution of the baryonic spin-isospin structure. In other words, the spin-isospin structure of the $\Delta$ baryon is formed by quarks that move in a coordinated manner. As the temperature increases, the baryon wave functions broaden, and the interactions between quarks weaken. These effects on the form factors versus temperature and momentum squared have been shown  for the axial vector Adler form factors  $C_{3}  (Q^{2}, T)$, $C_{4}  (Q^{2}, T)$, $C_{5}  (Q^{2}, T)$ and $C_{6}  (Q^{2}, T)$  in Figs.~\ref{7p} to \ref{9p}.

Taking the pion decay constant as $F = 0.087\text{ GeV}$ and fixing the soft-wall model parameters for the axial vector form factors to be:  
$$\kappa = -139.6, \; \tilde{\kappa} = 957, \;  \alpha_2 = 42, \;  \tilde{\alpha}_2 = 800, \;  \alpha_3 = 343, \;  \tilde{\alpha}_3 = -400, \;  \beta_2 = -20, \;  \tilde{\beta}_2 = -400 $$  
yields the axial vector $N \to \Delta$  transition form factors $C_{3}$, $C_{4}$, $C_{5}$ and  $C_{6}$ at zero momentum transfer and zero temperature as follows:
\begin{equation}
C_{3}(0,0) \approx 0.005, \quad C_{4}(0,0) \approx -0.3,\quad C_{5}(0,0) \approx 1.6, \quad   C_{6}(0,0) \approx 60.
\end{equation}
Note that these initial values of the form factor are fixed in the model using the parametrically deformed axial vector propagator with $\tau = -1$, rather than the standard axial propagator with $\tau = 1$. The reason for doing so will become apparent when we compare the form factor profiles using each of these propagator types. The above model parameters are fixed for both cases to enable a fair comparison. 

The initial form factor values, mentioned above, are those reported in the literature (see  \cite{geng2008nucleon}). Given the above parameters, our results in the thermal soft-wall model  predict profiles of the axial vector form factors at finite $Q^2$ and finite $T$. In some studies, the initial values above have been normalized differently \cite{alexandrou2007axial}, however, the shapes of the form factor profiles versus $Q^2$, remain consistent. A different normalization of the initial values of the form factors in the thermal soft-wall model can be achieved by tuning the model parameters mentioned above. We have checked that this does not significantly alter the shape of the profiles predicted by the soft-wall model. Thus, our results are modestly comparable to other existing methods, ranging from experimental measurements, to lattice QCD, to estimates from chiral perturbation theory \cite{unal2021axial, alexandrou2007axial, geng2008nucleon, ramalho2009nucleon, aliev2008light}. 

Our holographic model admits a non-vanishing $C_3(Q^2,T=0)$ form factor. In loop-based effective field theory approaches \cite{unal2021axial} this non-zero behavior is driven by higher-order loop contributions, whereas in our set-up it originates directly from the holographic structure of the action. This highlights the potential utility of considering $C_3(Q^2,T=0) \neq 0$ when higher-order corrections are taken into account.

The axial vector transition form factors $C_3(Q^2,T)$, $C_4(Q^2,T)$, $C_5(Q^2,T)$, and $C_6(Q^2,T)$ calculated in the thermal soft-wall model are shown as functions of $Q^2$ in Fig.~\ref{7p}(a), using  the standard axial vector propagator; and in Fig.~\ref{7p}(b), using the parametrically deformed axial propagator. The main difference between the two sets of results is the early $Q^2$ behavior. After about $Q^2 \approx 2$, both sets of profiles are nearly similar, and they both decrease monotonically with increasing $Q^2$.

Now let us compare the results in Fig.~\ref{7p} with the corresponding ones reported in \cite{geng2008nucleon}, \cite{alexandrou2007axial} and   \cite{ahn2009spin}. These are shown in Fig.~\ref{7plit}. The results of \cite{geng2008nucleon} are based on chiral perturbation theory, those of \cite{alexandrou2007axial} on lattice QCD methods, and \cite{ahn2009spin} refers to the holographic hard-wall model. Even between these different literature results, relative normalization of form factors, as well as scaling of the profiles, differ from author to author. What we will focus on is a comparison between the shape of the profiles. In this regard, notice that the profiles shown  in Fig.~\ref{7p}(b) closely resemble the shape of those in Fig.~\ref{7plit}(a), as well as the lattice QCD results in Fig.~\ref{7plit}(b). As for the profiles shown in Fig.~\ref{7p}(a), the resemblance (to \cite{geng2008nucleon} and \cite{alexandrou2007axial}) occurs only after $Q^2 >2$. On the other hand, the profiles in Fig.~\ref{7p}(a) seem closer to those obtained in the hard-wall model, possibly suggesting the relative similarity (and limitations) of the soft-wall model with the hard-wall model. Furthermore, the monotonic decrease in all the four form factors with increasing $Q^2$ is consistent with the findings of \cite{unal2021axial}.

\FloatBarrier

\FloatBarrier

Fig.~\ref{8pa} displays the temperature evolution of the axial vector $N \to \Delta$ transition form factors $C_3(Q^2,T)$, $C_4(Q^2,T)$, $C_5(Q^2,T)$, and $C_6(Q^2,T)$ at $Q^2 = 1, 2,$ and $3~\text{GeV}^2$. These results are based on the standard axial vector propagator. In Fig.~\ref{8p}, we show the same temperature profiles using the parametrically deformed axial propagator. The reason why the two sets of results are nearly identical is because the form factors, computed using the two kinds of propagators, are nearly identical after about $Q^2 \approx 2$.  

In all of the above, we find that $C_5(Q^2,T)$ and $C_6(Q^2,T)$  dominate the transition dynamics, attaining substantial magnitudes in the low-temperature regime.
On the other hand, $C_3(Q^2,T)$ and $C_4(Q^2,T)$ serve as subdominant corrections, suppressed by about an order of magnitude.  
As the temperature approaches the critical temperature $T_c \approx 0.22~\text{GeV}$, the axial gauge coupling in the 5D bulk weakens, thus driving all four form factors smoothly to zero ($C_i(Q^2,T) \to 0$). This universal drop indicates the thermal dissolution of hadrons prior to the deconfinement phase transition.

\FloatBarrier

Figure~\ref{9p} presents the $Q^2$ dependence of the axial vector $N \to \Delta$ transition form factors at fixed temperatures $T = 0.10, \; 0.15,$ and $0.20\text{ GeV}$, respectively. These figures involve the deformed axial propagator to show how an increase in temperature shifts the $Q^2$ form factor profiles. 
The form factors display distinct dynamical profiles:   $C_3(Q^2, T)$ in panel (a) and $C_5(Q^2, T)$ in panel (c) exhibit non-monotonic trends in the low-$Q^2$ region (the kinks seen in the  $C_5(Q^2, T)$ profiles here are likely numerical artifacts due to the resolution of the solver at those regions).  On the other hand, $C_4(Q^2, T)$ shown in panel (b) and $C_6(Q^2, T)$ shown in panel (d) vary monotonically, with $C_6(Q^2, T)$ exhibiting a potential power-law drop.    An increase in temperature suppresses the magnitude of all four form factors across the entire $Q^2$ spectrum, reflecting the thermal weakening of bulk couplings near the deconfinement boundary.  This temperature dependent suppression will also show up when using the standard axial propagator. More generally, temperature dependent suppression of form factors is a feature of the thermal soft-wall model, which also shows up for other hadron transitions \cite{nasibova2025pion}

\FloatBarrier

Table~\ref{tab2} shows all four axial vector form factor values based on the standard axial propagator, for various temperatures ($T$) and $Q^2$.   Table~\ref{tab3} shows the same, but for results computed using the deformed axial propagator. A comparison between the two tables reveals that the axial vector form factors obtained by the two propagators approach each other after about $Q^2 \approx 2$.

\begin{table}[htbp]  
\centering
\caption{Values of the $C_3, C_4, C_5,$ and $C_6$ form factors  as functions of $Q^2$ and $T$.} 
\vspace{0.5cm}   
\small
\setlength{\tabcolsep}{4pt}    
\begin{tabular}{cccccc c cccccc}  
\toprule
$Q^2$ & $T$ & $C_3$ & $C_4$ & $C_5$ & $C_6$ & &  $Q^2$ & $T$ & $C_3$ & $C_4$ & $C_5$ & $C_6$  \\  
\cmidrule{1-6}  \cmidrule{8-13} 
 0 & 0 & 2.53891 & -0.029202 & -0.232294 & 5.25448 &&  3.5 & 0 & 1.30132 & -0.0272923 & -16.8387 & 5.07309 \\ 
 0 & 0.1 & 2.0167 & -0.023722 & -0.165995 & 4.27675 && 3.5 & 0.1 & 0.953203 & -0.0189833 & -11.8561 & 3.53336 \\ 
 0 & 0.2 & 0.049440 & -0.000081 & -5.2 $\times 10^{-6}$ & 0.014214 &&  3.5 & 0.2 & 0.131941 & -0.00163946 & -1.09198 & 0.30348 \\ 
\addlinespace    
 0.5 & 0 & 2.79 & -0.094862 & -10.7884 & 17.807 && 4 & 0 & 1.21649 & -0.0229895 & -16.0507 & 4.26115 \\  
 0.5 & 0.1 & 2.04824 & -0.0781373 & -9.01324 & 14.7612 &&  4 & 0.1 & 0.887773 & -0.0158271 & -11.1876 & 2.93609 \\ 
 0.5 & 0.2 & 0.290248 & -0.018279 & -2.27196 & 3.58943 &&  4 & 0.2 & 0.118747 & -0.00130453 & -0.985284 & 0.240156 \\  
\addlinespace   
 1 & 0 & 2.12759 & -0.0802952 & -16.4526 & 15.1196 && 4.5 & 0 & 1.14103 & -0.0195929 & -15.2708 & 3.6216 \\  
 1 & 0.1 & 1.54285 & -0.0625594 & -12.987 & 11.8441 &&   4.5 & 0.1 & 0.829582 & -0.0133719 & -10.5528 & 2.47274 \\  
 1 & 0.2 & 0.241968 & -0.0101679 & -2.25391 & 1.97756 &&  4.5 & 0.2 & 0.107754 & -0.00105984 & -0.894959 & 0.194153 \\  
\addlinespace   
 1.5 & 0 & 1.82777 & -0.0631577 & -18.3083 & 11.8748 &&   5 & 0 & 1.07252 & -0.016906 & -14.5548 & 3.11707 \\  
 1.5 & 0.1 & 1.33795 & -0.0473765 & -13.9004 & 8.94649 &&   5 & 0.1 & 0.776899 & -0.0114577 & -9.98925 & 2.11274 \\  
 1.5 & 0.2 & 0.21755 & -0.00607945 & -1.88884 & 1.16565 &&  5 & 0.2 & 0.0984744 & -0.000882161 & -0.824089 & 0.16098 \\  
\addlinespace   
 2 & 0 & 1.64511 & -0.0499517 & -18.6127 & 9.36746 && 5.5 & 0 & 1.01097 & -0.0147183 & -13.8725 & 2.70714 \\  
 2 & 0.1 & 1.21061 & -0.0364682 & -13.7498 & 6.86217 &&  5.5 & 0.1 & 0.7298 & -0.00991418 & -9.46354 & 1.82317 \\  
 2 & 0.2 & 0.190984 & -0.00400973 & -1.60037 & 0.759795 &&  5.5 & 0.2 & 0.0905848 & -0.000744058 & -0.761767 & 0.135295 \\  
\addlinespace   
 2.5 & 0 & 1.50913 & -0.0401686 & -18.2528 & 7.51091 && 6 & 0 & 0.955195 & -0.0129244 & -13.2385 & 2.37177 \\  
 2.5 & 0.1 & 1.11121 & -0.0287395 & -13.2153 & 5.3878 && 6 & 0.1 & 0.687346 & -0.00866009 & -8.98409 & 1.58852 \\  
 2.5 & 0.2 & 0.167423 & -0.00284368 & -1.38759 & 0.533856 &&  6 & 0.2 & 0.0838046 & -0.000636119 & -0.708345 & 0.115306 \\  
\addlinespace   
 3 & 0 & 1.3974 & -0.0328571 & -17.6041 & 6.12549  & &  &&&&& \\ 
 3 & 0.1 & 1.02687 & -0.0231431 & -12.5498 & 4.32283  & &  &&&&& \\ 
 3 & 0.2 & 0.147916 & -0.00212088 & -1.22375 & 0.395148  & &  &&&&& \\   
\bottomrule
\end{tabular}
\label{tab2}  
\end{table}

\clearpage

\begin{table}[htbp]
\centering
\caption{Values of the $C_3, C_4, C_5,$ and $C_6$ form factors  as functions of $Q^2$ and $T$.}  
\vspace{0.5cm} 
\small
\setlength{\tabcolsep}{4pt}
\begin{tabular}{cccccc c cccccc}
\toprule
$Q^2$ & $T$ & $C_3$ & $C_4$ & $C_5$ & $C_6$ & &  $Q^2$ & $T$ & $C_3$ & $C_4$ & $C_5$ & $C_6$  \\  
\cmidrule{1-6}  \cmidrule{8-13} 
0 & 0 & 0.00547801 & -0.311771 & 1.62171 & 60.1398  & &  3.5 & 0 & 1.27921 & -0.0251959 & -12.3047 & 4.67589  \\
 0 & 0.1 & -0.637097 & -0.264305 & -4.67414 & 51.4751 & &  3.5 & 0.1 & 0.938909 & -0.0174407 & -9.06756 & 3.23981 \\
 0 & 0.2 & -1.0413 & -0.0917138 & -8.37716 & 19.0594 & &   3.5 & 0.2 & 0.125451 & -0.00192555 & -1.32644 & 0.363054 \\  
\addlinespace  
 0.5 & 0 & 1.24953 & -0.154054 & -8.62139 & 29.4864 & &  4 & 0 & 1.19814 & -0.0211071 & -11.953 & 3.90437 \\
 0.5 & 0.1 & 0.860547 & -0.118767 & -9.3432 & 22.8813 & & 4 & 0.1 & 0.874655 & -0.0145275 & -8.71059 & 2.68914 \\ 
 0.5 & 0.2 & 0.206615 & -0.0195298 & -2.89185 & 3.87776 & &  4 & 0.2 & 0.115527 & -0.00153411 & -1.16954 & 0.28694 \\ 
\addlinespace  
 1 & 0 & 1.56874 & -0.095329 & -11.488 & 18.1146 & &  4.5 & 0 & 1.12372 & -0.0179896 & -11.6147 & 3.31819 \\ 
 1 & 0.1 & 1.1803 & -0.070498 & -10.143 & 13.4571 & &   4.5 & 0.1 & 0.81648 & -0.0123302 & -8.39419 & 2.27546 \\ 
 1 & 0.2 & 0.244845 & -0.00903215 & -1.97801 & 1.75008 & &  4.5 & 0.2 & 0.106818 & -0.00118497 & -1.00332 & 0.219223 \\  
\addlinespace 
 1.5 & 0 & 1.5957 & -0.0664538 & -12.6399 & 12.552 & & 5 & 0 & 1.05648 & -0.0154991 & -11.2351 & 2.85109 \\
 1.5 & 0.1 & 1.20365 & -0.0480422 & -10.33 & 9.10464 & &  5 & 0.1 & 0.764557 & -0.0105781 & -8.05693 & 1.94647 \\ 
 1.5 & 0.2 & 0.214452 & -0.00556802 & -1.6887 & 1.06544 & &  5 & 0.2 & 0.0981459 & -0.00096542 & -0.903068 & 0.177513 \\  
\addlinespace  
 2 & 0 & 1.53888 & -0.0493847 & -12.9564 & 9.27916 & &   5.5 & 0 & 0.995389 & -0.0135223 & -10.8895 & 2.48146 \\  
 2 & 0.1 & 1.15325 & -0.0351531 & -10.1529 & 6.62146 & &   5.5 & 0.1 & 0.717828 & -0.0092003 & -7.76363 & 1.68868 \\ 
 2 & 0.2 & 0.185612 & -0.00385085 & -1.50897 & 0.730636 & & 5.5 & 0.2 & 0.0905674 & -0.000799699 & -0.818997 & 0.146248 \\  
\addlinespace  
 2.5 & 0 & 1.45548 & -0.0383055 & -12.8785 & 7.16411 & &  6 & 0 & 0.940051 & -0.0119114 & -10.5573 & 2.18105 \\  
 2.5 & 0.1 & 1.08258 & -0.0269492 & -9.8187 & 5.0494 & &   6 & 0.1 & 0.67586 & -0.00808366 & -7.49066 & 1.48035 \\  
 2.5 & 0.2 & 0.16219 & -0.00284835 & -1.37479 & 0.537039 & &  6 & 0.2 & 0.0839155 & -0.000674859 & -0.751355 & 0.122877 \\ 
\addlinespace  
 3 & 0 & 1.36616 & -0.0306766 & -12.6249 & 5.71374 & &  &&&&& \\
 3 & 0.1 & 1.00902 & -0.0213857 & -9.43904 & 3.98857 & &  &&&&& \\
 3 & 0.2 & 0.143112 & -0.00222923 & -1.28653 & 0.418664 & &  &&&&& \\  
\bottomrule
\end{tabular}
\label{tab3}
\end{table}

\section{Conclusions and Discussion}

\label{con}
In this work, we have carried out a comprehensive investigation of the momentum transfer and temperature dependence of the electromagnetic ($G_{M_1}$, $G_{E_2}$, $G_{C_2}$) and axial vector ($C_3$, $C_4$, $C_5$, $C_6$) multipole form factors for the $N \rightarrow \Delta$ transition using the thermal AdS/QCD soft-wall model with regularized propagators.   
In the course of computing these form factors from the soft-wall model, the main stumbling block that had to be addressed was the recently reported divergence in the axial vector bulk-to-boundary propagator \cite{leutgeb2026divergences}. The transverse axial field EOM includes a VEV term originating from the bulk scalar. This term depends on the holographic coordinate, leading to a divergence in the IR limit, which, if taken at face-value, is pathological for the soft-wall model. In this work, we have addressed this problem concerning the axial vector propagator. We have shown that a regularization in the form of a momentum-dependent cut-off can be achieved via an interior-constraint approach to the boundary-value problem. This modified propagator is finite. As noted above, this approach of modifying the propagator is only a phenomenological fix. As we comment below, a substantial modification of the soft-wall model would be required to ensure finiteness of the propagator from first principles. 

On the other hand, the practical utility of implementing the above regularization within the soft-wall model is demonstrated through our numerical estimates of the EM and AV form factors of the $N \to \Delta$ transition. The moderately reasonable comparison of these profiles with experimental and lattice QCD results at zero temperature suggests the practical use of having an IR cut-off.  There are two versions of the modified axial propagator that we have tested: first, there is the standard axial propagator with $\tau = 1$ in Eq.~\ref{ax2}, and then we also tested the parametrically deformed propagator with $\tau = -1$. With the standard propagator, our results match the experimental/lattice results only for $Q^2 > 2 \; GeV^2$, but with the deformed propagator, we found a remarkably close fit with the shape of the form factor profiles in the literature for the entire range of $Q^2$. It remains to be seen whether this observation carries a deeper theoretical significance. On the other hand, the results obtained using the standard axial propagator are comparable to those of the AdS/QCD hard-wall model (in \cite{ahn2009spin}), thus reflecting the similarity and limitations of both, the hard-wall and soft-wall models.

We then proceed further and investigate the temperature dependence of the same $N \to \Delta$ form factor profiles within the thermal soft-wall model. We find that all seven $N\to\Delta$ transition form factors ($G_{M_1}$, $G_{E_2}$, $G_{C_2}$, $C_3$–$C_6$) are suppressed as $T \to T_c$. This decrease in the thermal profiles of the dominant $G_{M_1}$ and $C_5, \; C_6$ form factors indicates a weakening of the $N \to \Delta$ transition with increasing temperature, which is consistent with thermal screening of the respective currents. The quadrupole form factors $G_{E_2}$ and $G_{C_2}$, and the sub-leading transverse ones $C_3$ and  $C_4$, reflect the departure from spherical symmetry. Namely, their reduction may be interpreted as a tendency toward a more symmetric transition. The decrease of $C_6$ with increase in temperature is  governed by the pion pole, consistent with the weakening of the long-range pion cloud. Overall, the interaction appears to lose both its deformation and its pion-cloud component as the temperature increases. This is one qualitative interpretation of the thermal effects we have  observed here. The same suppression may also be related to thermal broadening and dissociation of hadron states, and the two effects may occur simultaneously. Separating them quantitatively is left for future work.

All in all, the thermal AdS/QCD soft-wall model, supplemented by the proposed interior-constraint method, provides an empirically reliable framework for investigating hadron transition form factors in hot QCD matter. Beyond resolving the asymptotic divergence in the axial vector sector, the present approach offers a practical tool for studying a broad class of electromagnetic and axial vector processes involving excited baryons under extreme conditions. Presumably, the formalism can be systematically extended to finite baryon density, external electromagnetic fields, and non-equilibrium environments, providing valuable phenomenological predictions for future heavy-ion collision experiments, lattice QCD simulations, and the exploration of strongly interacting matter in extreme astrophysical phenomena (for instance, holographic models of neutron stars or multi-center black holes \cite{jarvinen2022holographic, Arsiwalla:2010thesis, Arsiwalla:2011}).  

For future work, it may be worth examining other systematic modifications to the soft-wall model that address all the concerns raised in \cite{leutgeb2026divergences}. Even though the modified  propagators we have proposed here work for empirical purposes, it would be insightful to understand how IR completeness might be realizable within the core formalism of the soft-wall model (possibly, involving modifications to the Lagrangian or couplings). Unlike the hard-wall model, which has a rigid cut-off, a $Q^2$ (and $T$) dependent cut-off, as the one we have here, is more natural, as it follows directly from the interior boundary-value problem. All of this suggests the need for a hybrid holographic model, which retains the main advantages of the thermal soft-wall model (namely, fitting the Regge trajectories, plus the empirical form factor estimates), while incorporating features of other bottom-up approaches, such as V-QCD \cite{jarvinen2012holographic}, or even geometric features from top-down thermal holographic QCD, such as in \cite{chen2013ultraviolet}. This is ongoing research, which we shall report in a forthcoming publication.

\bibliographystyle{unsrt}
\bibliography{references}
\end{document}